\documentclass[useAMS, usenatbib]{mn2e}
\usepackage{amsmath}
\usepackage{amsfonts}
\usepackage{amssymb}
\usepackage{makeidx}
\usepackage{graphicx}
\usepackage{url}
\usepackage{comment}
\usepackage{xcolor}
\usepackage{aas_macros} 
\usepackage{comment}
\usepackage{longtable}
\usepackage{multirow}
\definecolor{darkblue}{rgb}{0,0,.5}
\usepackage[dvipdfm,colorlinks,citecolor=darkblue,linkcolor=blue,urlcolor=blue]{hyperref}
\usepackage{times}

\title{The observed radio/gamma-ray emission correlation for blazars with the \textit{Fermi}-LAT and the RATAN-600 data}
\author[T.~Mufakharov]
 {T.~Mufakharov,$^1$\thanks{E-mail:
timur.mufakharov@gmail.com;}
  M.~Mingaliev,$^{1,2}$
  Yu.~Sotnikova,$^1$
   Ya.~Naiden$^1$ and
     A.~Erkenov$^1$\\
  $^1$Special Astrophysical Observatory of RAS, Nizhnij Arkhyz, 369167 Russia\\
  $^2$Kazan Federal University, 18 Kremlyovskaya St., Kazan, 420008, Russia}
\date{Accepted 2015 April 6.  Received 2015 March 18; in original form 2014 September 25}

\pagerange{\pageref{firstpage}--\pageref{fig5}} \pubyear{2015}

\def\LaTeX{L\kern-.36em\raise.3ex\hbox{a}\kern-.15em
    T\kern-.1667em\lower.7ex\hbox{E}\kern-.125emX}
\begin{document}

\label{firstpage}

\maketitle

\begin{abstract}
We study the correlation between gamma-ray and radio band radiation for 123 blazars,
 using the \textit{Fermi}-LAT first source catalog (1FGL) and the RATAN-600 data obtained at the same period of time (within a few months).
We found an apparent positive correlation for BL Lac and flat-spectrum radio quasar (FSRQ) sources from our sample through testing the value of the Pearson product-moment correlation coefficient.
The BL Lac objects show higher values of the correlation coefficient than FSRQs at all frequencies, except 21.7 GHz,
 and at all bands, except $10-100$ GeV, typically at high confidence level ($>$ 99 \%).
At higher gamma-ray energies the correlation weakens and even becomes negative for BL Lacs and FSRQs.
For BL Lac blazars, the correlation of the fluxes appeared to be more sensitive to the considered gamma-ray energy band, than to the frequency, while for FSRQ sources the correlation  changed notably both with the considered radio frequency and gamma-ray energy band.
We used a data randomization method to quantify the significance of the computed correlation coefficients.
We find that the statistical significance of the correlations we obtained between the flux densities at all frequencies 
and the photon flux in all gamma-ray bands below 3 GeV is high for BL Lacs (chance probability $\sim 10^{-3} - 10^{-7}$).
The correlation coefficient is high and significant for the $0.1-0.3$ GeV band and low and insignificant for the $10-100$ GeV band
for both types of blazars for all considered frequencies.
\end{abstract}


\section{Introduction}
Blazars are the most extreme class of active galactic nuclei (AGNs)
characterized by a prominent jet pointing within a few degrees of our line of sight \citep{1995PASP..107..803U}. 
They are known to be highly variable at 
gamma-ray wavelengths, as well as in the radio band, on time scales of days to months \citep{1995ApJ...440..525V}.
Observationally they are divided into two main classes: the BL Lac objects (BL Lacs) with an almost featureless spectrum and the flat-spectrum radio quasars (FSRQs) with strong broad emission lines in their spectrum \citep{1995PASP..107..803U}.

The spectral energy distribution (SED) of blazars
is characterized by two broad features.
The first one, peaking at lower energy, is generally explained
in terms of synchrotron emission;
 the second feature, peaking at higher energies, 
is likely due to inverse Compton radiation \citep{1996ApJ...463..444S}.
Most of the observed radio to optical (and in some cases X-ray)
emission from the blazars is due to synchrotron radiation in the jet \citep{1981Natur.293..714B,1982ApJ...253...38U,1988AJ.....95..307I,1998ASPC..144...25M}.
Synchrotron emission is produced by relativistic electrons,
moving in a magnetic field (\citealt{1979rpa..book.....R}).
Inverse Compton photons originate from the interaction
of the energetic electrons with seed photons.
These seed photons could be produced by synchrotron emission,
via synchrotron self-Compton (SSC) radiation (e.g., \citealt{1981ApJ...243..700K,1985ApJ...298..114M}),
or they might originate from some external source
 - in this case it is external inverse Compton radiation (e.g., \citealt{1994ApJ...421..153S,1995ApJ...441...79B}).
The gamma-ray photons could originate
through the SSC mechanism, and then one might
expect significant correlation 
between gamma-ray and radio emission because of the same origin
for the radio and gamma-ray photons.
On the other hand, if there is no reliable proof
for such connection, that would support the theory 
of the independent origin of these emissions.

Since the majority of AGNs identified with gamma-ray sources are also bright radio sources - about half of the 1400 gamma-ray sources from the first \textit{Fermi}-LAT
catalogue \citep{2010ApJ...715..429A} were identified with radio sources - this provides the motivation to search for a correlation between radiation in gamma-ray and radio bands. By investigating such correlation one can study time delays between the different events in the gamma-ray and radio band light curves, the physical processes and the characteristics of the radiation in the AGN jets. The presence or absence of correlation also can help to more accurately determine the parameters of the models for the structure and processes in AGNs.

The first dedicated studies of the gamma-ray-radio emission correlation in blazars were based on EGRET data (e.g., \citealt{1993ApJ...410L..71S,1993MNRAS.260L..21P}). However, these results remain uncertain due to the use of observational data obtained non-simultaneously, and also because samples were flux limited \citep{1997A&A...320...33M,2007ApJ...671.1355T}.

The search for significant correlation between gamma-ray and radio emission continued, when \textit{Fermi}-LAT telescope data became available.
Using gamma-ray data from EGRET and \textit{Fermi} surveys, \citet{2010MNRAS.407..791G} noted that flux could change up to three times during the year.
When average flux values are used for analysis, short-term variability (from one to several days or weeks) did not greatly affect the variability averaged over the year. The $F\gamma$ -- $F_{r}$ correlation has been studied by \citet{2010MNRAS.407..791G} between the gamma-ray flux above 100 MeV (using 1FGL catalogue; \citealt{2010ApJS..188..405A}) and the 20 GHz flux density (using ATCA survey; \citealt{2010MNRAS.402.2403M}).
A statistically significant (more than 3$\sigma$) correlation was found 
  both for the population of BL Lac and FSRQ sources. Also they considered selection effects (sensitivity limits for radio and gamma-ray telescopes) and the likelihood that some radio sources were not detected due to their variability in the gamma-ray band.

\citet{2009ApJ...696L..17K} also investigated the correlation between radio emission and gamma-ray flux (using \textit{Fermi}-LAT data after the first three months of observations) for 135 AGNs. The non-parametric Kendall tau test confirmed a positive correlation at a confidence level greater than 99.9 \%
between the gamma-ray flux (for the 100 MeV-1 GeV energy band) and radio flux density (measured by the Very Long Baseline Array  at 15 GHz within
several months of the \textit{Fermi}-LAT observations). The same analysis for the second \textit{Fermi}-LAT energy band,
$1-100$ GeV, also showed a positive correlation, but at a lower confidence level - 86 \%.

\citet{2011ApJ...741...30A} performed a detailed statistical analysis of the correlation between the radio and gamma-ray emission of the AGNs detected by \textit{Fermi}-LAT in its first year of operation.
For the radio band they used archival data at 8 GHz for 599 sources and concurrent measurements at 15 GHz for 199 sources, provided by the Owens Valley
Radio Observatory monitoring programme \citep{2011ApJS..194...29R}. 
One distinctive feature of that work was the study of not only the \textit{apparent}, but also the \textit{intrinsic} strength of the correlation, exploiting a new statistical method by \citet{2012ApJ...751..149P}.
They found that the statistical significance of a positive correlation between the centimetre radio and the broadband (E $>$ 100 MeV) gamma-ray energy flux is very high for both flat spectrum radio quasars and BL Lac objects from their AGN sample.

Moreover, the correlation between high frequency radio emission (at 37 GHz)
and gamma-ray emission (100 MeV-100 GeV \textit{Fermi}-LAT data) for 249 northern AGNs was studied by \citet{2011A&A...535A..69N}.
They also found a significant correlation between both the flux densities and luminosities in the gamma-ray and radio bands
and suggested that the gamma radiation is
produced co-spatially with the 37 GHz emission, i.e. in the jet.

On the basis of the work,\ considered above, it can be concluded that homogeneous (derived from one instrument) and simultaneous observational data in the radio and gamma-ray bands are essential for detection a possible correlation in radiation from blazars.

In this paper we present a statistical analysis of the correlation between
radio and gamma-ray emission for a sample of 123 blazars using quasi-simultaneous
observational data from the \textit{Fermi}-LAT and the RATAN-600 telescopes.
The correlation analysis was performed using the flux densities from five radio bands,
21.7, 11.2, 7.7, 4.8 and 2.3 GHz, for the first time.

\section{Sample sources and observation}
For the correlation analysis we cross-match the first \textit{Fermi}-LAT catalogue (1FGL) with the RATAN-600 observations. 
The 1FGL catalogue is available in the Vizier database\footnote{http://vizier.u-strasbg.fr/} and its description is given in \citet{2010ApJ...715..429A}.
The catalogue includes observational data obtained during the period from 2008 August 4 to 2009 July 4 for 1451 AGNs
and contains information for five gamma-ray bands: $0.1-0.3$, $0.3-1$, $1-3$, $3-10$ and $10-100$ GeV. Gamma-ray flux units are photons per second per square centimetre (ph cm$^{-2}$ s$^{-1}$). The counts in each band are averaged over 11 months.

About 300 AGNs were observed with the RATAN-600 radio telescope
within a few months of the \textit{Fermi}-LAT observations, as part of various programmes.
For the analysis we used flux densities measured at five frequencies (2.3, 4.8, 7.7, 11.2 and 21.7 GHz) in 2008 November and 2009 April.
Each source was observed 5 to 10 times during this period.
The experimental data were processed with the modules of the FADPS
(Flexible Astronomical Data Processing System) standard reduction
package by \citet{1997ASPC..125...46V}. The processing
methods are described in the paper by \citet{2012A&A...544A..25M}.
The following 12 flux density calibrators (standard and
RATAN's traditional ones) were used to
get the coefficients of antenna elevation: 3C48,
3C138, 3C147, 3C161, 3C286, 3C295, 3C309.1, NGC7027,
J0237$-$23, J1154$-$35, J1347$+$12 and J0410$+$76.
Measurements of some calibrators were corrected for angular
size and linear polarization, following the data summarized in \citet{1994A&A...284..331O}
and \citet{1980A&AS...39..379T}, respectively.
The detection limit for the RATAN-600 single sector
is approximately 8~mJy (integration time is about 3 s) under good weather conditions
at the frequency of 4.8 GHz and at an average antenna elevation ($\delta\sim$ $42^{\circ}$).
The detection limits at other frequencies
are presented in Table~\ref{tab:radiometers} along with the other parameters of radiometers:
$f_{0}$ -- central frequency, $\Delta$$f_{0}$ -- bandwidth, $\Delta$$F$ -- flux density detection limit per beam, and
BW -- beam width [full width at half-maximum (FWHM) in right ascension (RA)] at average antenna elevation ($\delta\sim$ $42^{\circ}$). 
Beam width in declination is three to five times worse than in RA.
These values depend on the atmospheric extinction instability and the effective
area at the antenna elevation $H$ (from $10^{\circ}$ up to $90^{\circ}$ above the horizont)
at the corresponding frequencies.
Radio data that are used in this paper are published in RATAN-600 BL Lacs database\footnote{http://www.sao.ru/blcat/}\citep{2014A&A...572A..59M} and in \citet{2012A&A...544A..25M}.
Standard errors in determining the flux density for these data are
7-8 \% at 2.3 GHz and 4-5 \% at other frequencies. 
Almost all of the sources had relatively strong flux levels
 at radio frequencies with a signal-to-noise ratio $S/N \geq 4$.

\begin{table}
\caption{RATAN-600 continuum radiometers.}
\label{tab:radiometers}
\centering
\begin{tabular}{cccc}
\hline
$f_{0}$  & $\Delta$$f_{0}$ & $\Delta$$F$  & BW      \\
(GHz)    & (GHz)            & (mJy beam$^{-1}$)    &  (arcsec) \\
\hline
21.7 &  2.5  &  70  &  11 \\
11.2 &  1.4  &  20  &  16 \\
7.7  &  1.0  &  25  &  22 \\
4.8  &  0.9  &  8   &  36 \\
2.3  &  0.4  &  30  &  80 \\
\hline
\end{tabular}

\medskip
Column designation:
Col.~1~-- central frequency,
Col.~2~-- bandwidth,
Col.~3~-- flux density detection limit per beam, and
Col.~4~-- beam width (FWHM in RA).
\end{table}

\begin{figure}
\center{\includegraphics[width=1\linewidth]{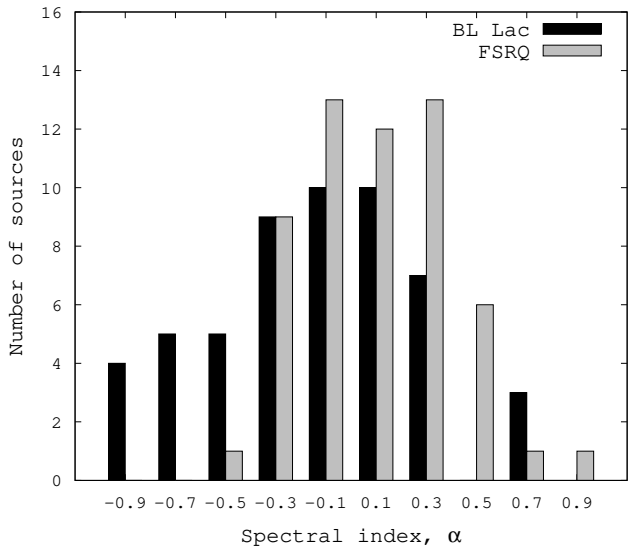}} \\ 
\caption{Distribution of the spectral indices, measured between 2.3 and 7.7 GHz for BL Lacs (shown with black) and FSRQs (shown with grey)}
\label{sp_indices}
\end{figure}

\begin{figure}
\center{\includegraphics[width=1\linewidth]{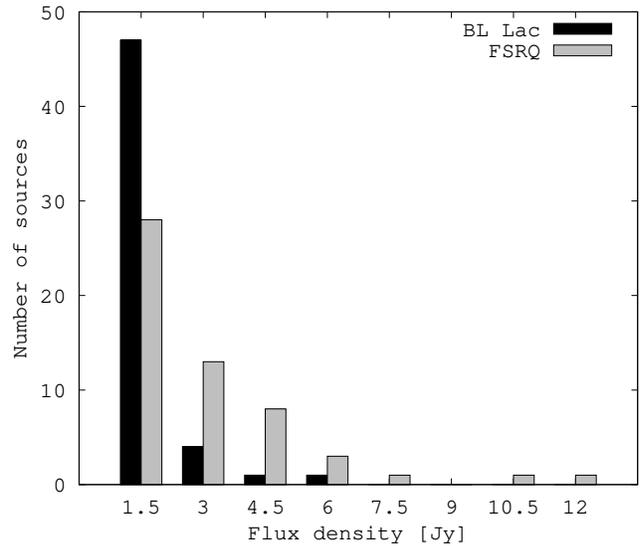}} \\ 
\caption{Distribution of the flux densities at 4.8 GHz, measured with the RATAN-600, for BL Lacs (shown with black) and FSRQs (shown with grey). We note that flux density does not reach 1 Jy for most of the BL Lacs}
\label{fig0}
\end{figure}

After cross-matching we had 123 AGNs for which quasi-simultaneous \textit{Fermi}-LAT and RATAN-600 data were available.
They make up the final list for the further analysis. In Table \ref{table} we present: source name [from the NASA/IPAC (NED)\footnote{http://ned.ipac.caltech.edu/} or from the Roma-BZCAT catalogue\footnote{http://www.asdc.asi.it/bzcat/} \citep{2009A&A...495..691M}] (column 1), blazar type (Roma-BZCAT) (column 2), and redshift value (NED) (column 3). 
There are different sub-classes of blazars presented in this sample: 53 -- BL Lacs, 6 -- BL Lac candidates, 8 -- blazars of uncertain type, 56 -- FSRQs.
Almost all FSRQs and most of the BL Lacs have a flat radio spectrum ($|\alpha| \le 0.5$), Figure \ref{sp_indices} shows the spectral index distribution for FSRQs and BL Lacs from our sample.
Figure \ref{fig0} shows the distribution of flux densities, measured at 4.8 GHz, for BL Lacs and FSRQs (we exclude 3C 273 ($F > 39$ Jy) from this distribution for clarity of the display).  The flux density does not reach 1 Jy for most of the BL Lacs at all frequencies. Note that FSRQs, observed with the RATAN-600, are limited by flux density ($\geq$ 1 Jy). Moreover the representation of BL Lacs in our sample is incomplete, since they were observed partly in different programmes.
In Table \ref{fluxes}, we list the radio flux densities at different frequencies used in this work, along with the spectral index values ($\alpha$), measured between 2.3 and 7.7 GHz, but for two sources measured between 4.8 and 7.7 GHz (due to the absence of observations at 2.3 GHz for them).

\section{Results}
\subsection{Flux density correlation}
With the broadband observational data available for radio and gamma-ray bands, we compared correlation coefficients for a relatively large and approximately equal number of sources from two blazar sub-classes (BL Lacs and FSRQs). We checked the possibility of the existence of the flux density correlation measured at five radio bands with photon flux in five gamma-ray bands. 
We calculate the Pearson product-moment correlation coefficient $r$ between radio and gamma-ray flux for the objects of our sample. In Table \ref{table_corr} we present correlation coefficients for five radio and five gamma-ray bands, along with the number of sources (N) with available radio data at each frequency (we have data for the full sample in gamma-rays) and typical confidence level values (CL) for calculated correlation coefficients for each considered band.
Figures \ref{fig1} - \ref{fig5} are $F\gamma$ -- $F_{r}$ plots.
The sub-samples are plotted with different symbols: BL Lac - empty circles, FSRQ - filled triangles, BL Lac candidates - empty squares, blazar of uncertain type - filled squares.

\begin{figure}
\center{\includegraphics[width=1\linewidth]{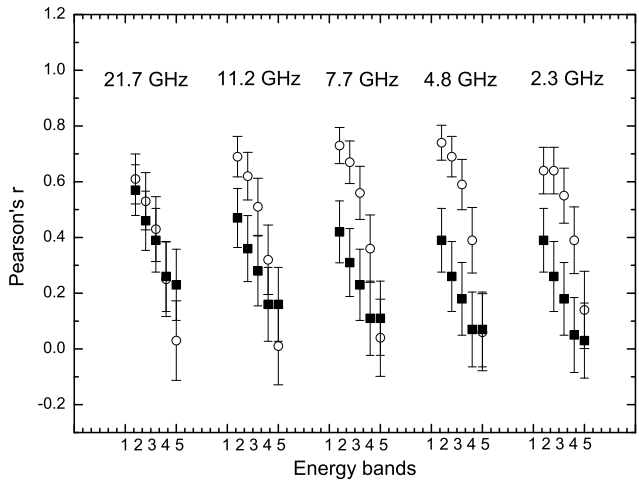}} \\ 
\caption{Pearson's $r$ distribution for the radio flux density and gamma-ray photon flux as a function of the
energy bands shown for five frequencies. Correlation coefficients shown with empty circles for BL Lacs and with filled squares for FSRQs.
Five energy bands are marked with numbers 1--5 (0.1-0.3, 0.3-1, 1-3, 3-10, and 10-100 GeV, respectively) in a horizontal axis for each frequency: 21.7, 11.2, 7.7, 4.8 and 2.3 GHz. Plotted values of the Pearson's $r$ are from Table \ref{table_corr}}
\label{Cor-Energy}
\end{figure}

\begin{figure*}
\centering
\begin{tabular}{ccc}
\begin{minipage}{0.32\linewidth}
\center{\includegraphics[width=\textwidth]{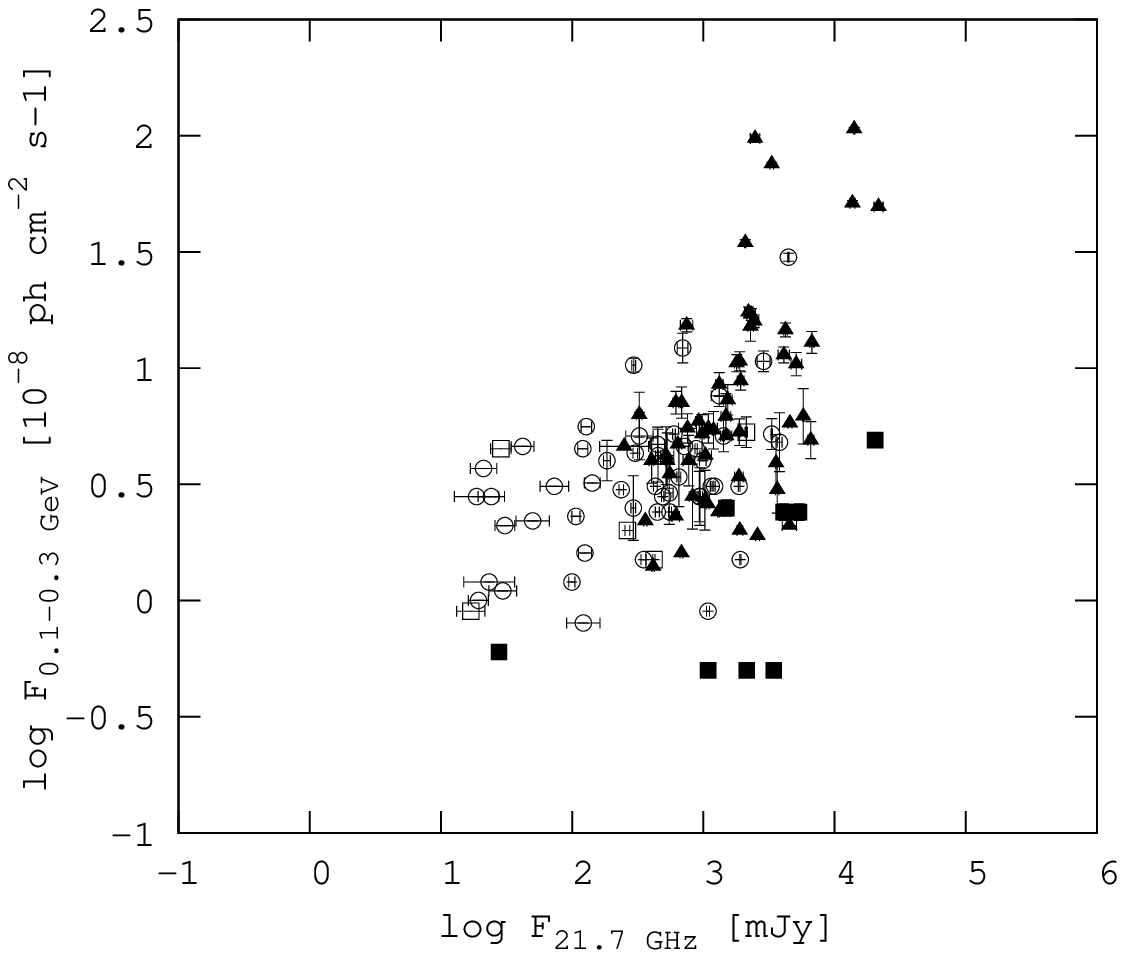}} \\ 
\end{minipage}
\hfill
\begin{minipage}{0.32\linewidth}
\center{\includegraphics[width=\textwidth]{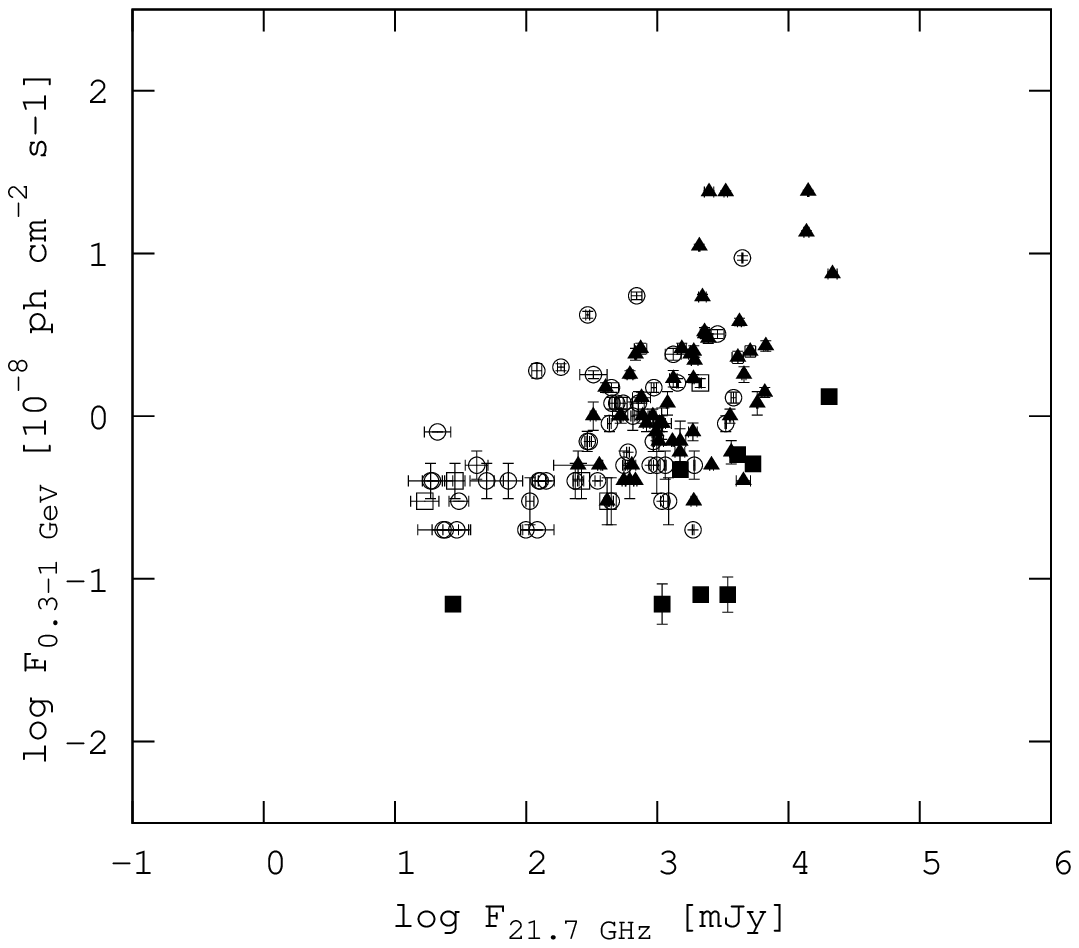}} \\ 
\end{minipage}
\hfill
\begin{minipage}{0.32\linewidth}
\center{\includegraphics[width=\textwidth]{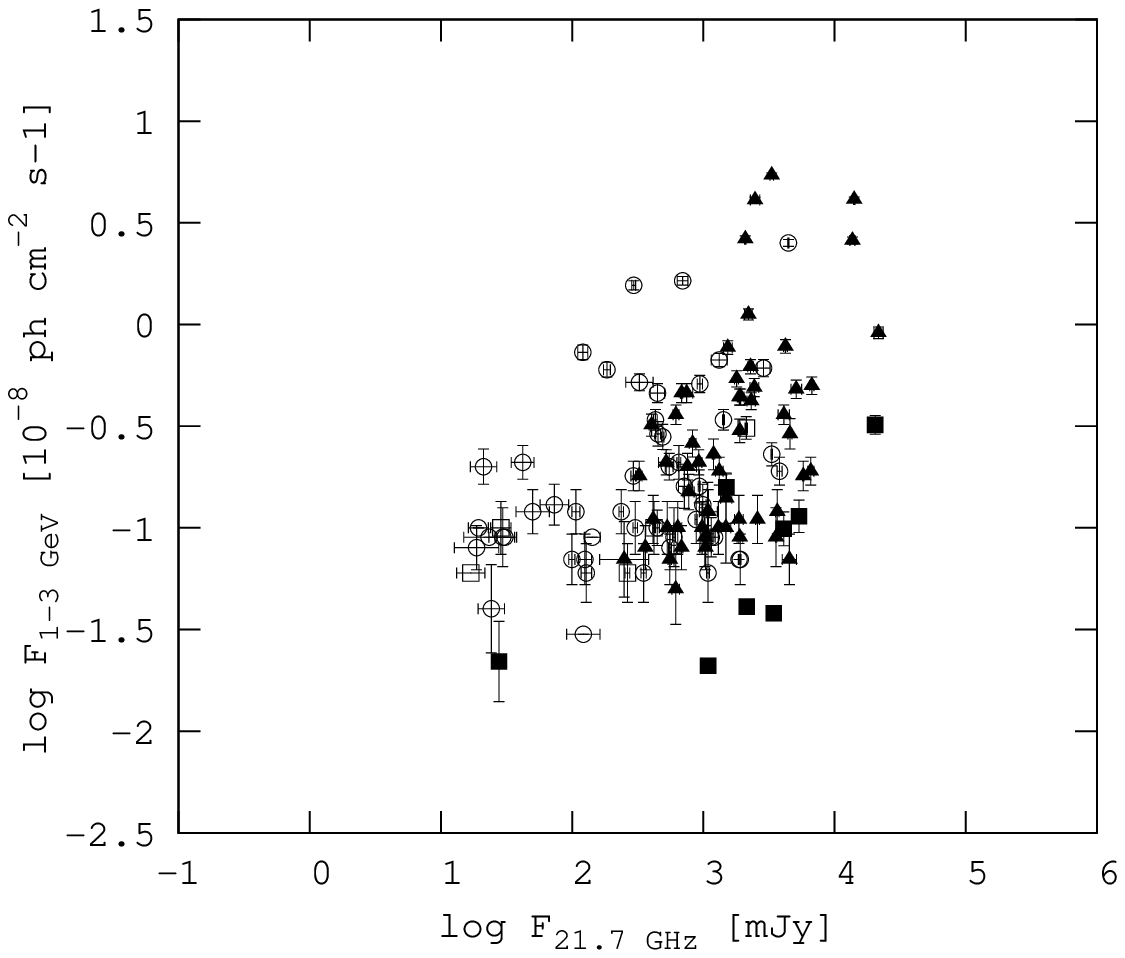}} \\ 
\end{minipage}
\\
\\
\begin{minipage}{0.32\linewidth}
\center{\includegraphics[width=1\linewidth]{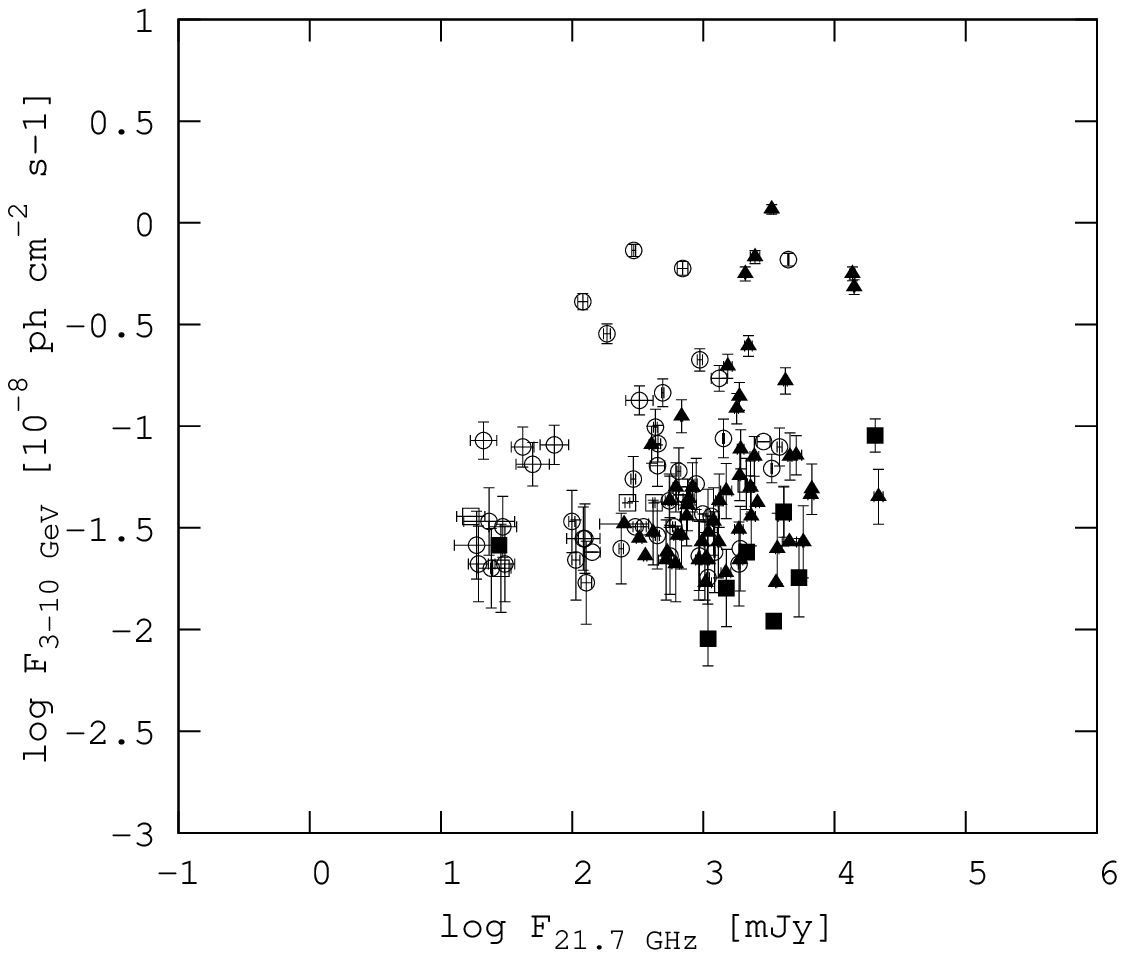}} \\   
\end{minipage}
\begin{minipage}{0.32\linewidth}
\center{\includegraphics[width=1\linewidth]{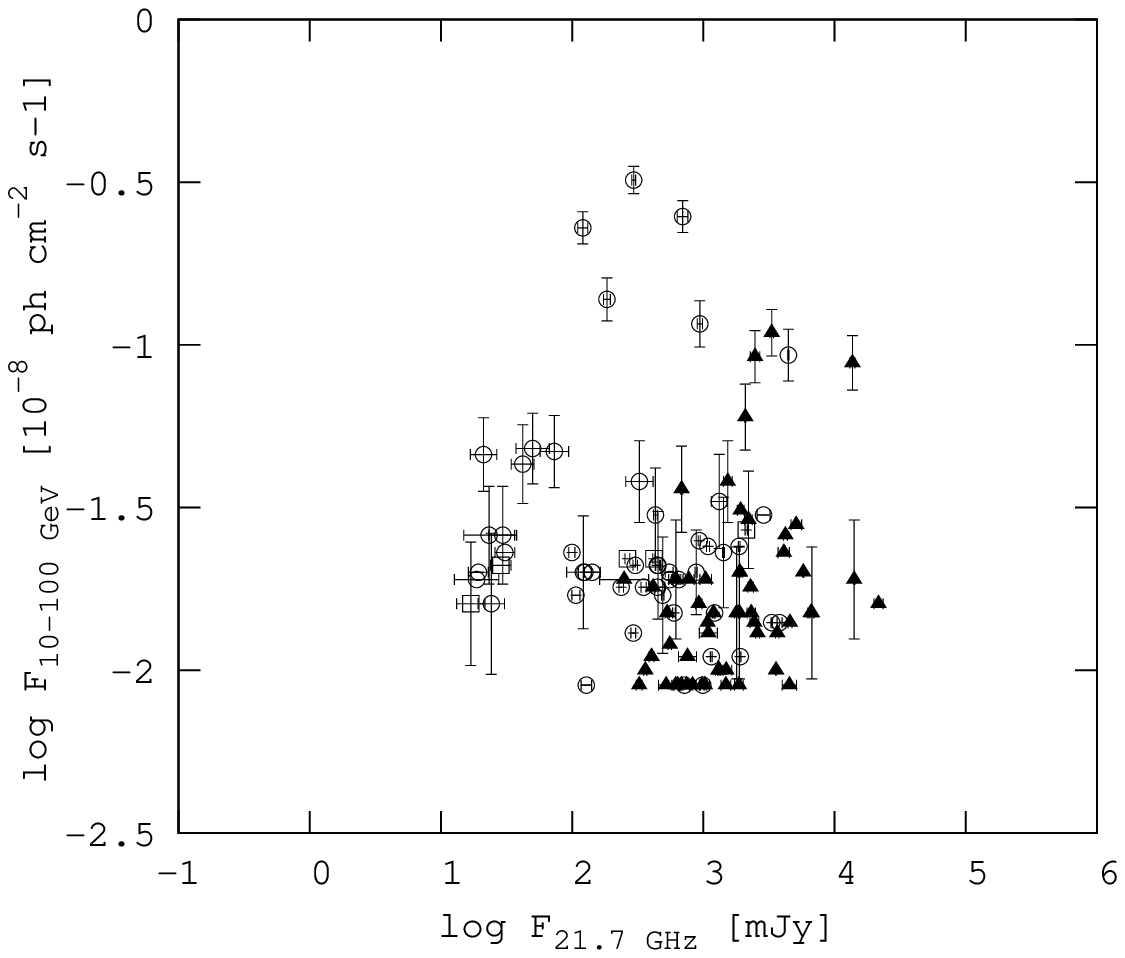}} \\ 
\end{minipage}
\end{tabular}
\caption{Broadband gamma-ray photon flux versus 21.7 GHz flux density. BL Lacs shown with empty circles, FSRQ - filled triangles, BL Lac candidates - empty squares, Blazars of uncertain type - filled squares.}
\label{fig1}
\end{figure*}

\begin{figure*}
\centering
\begin{tabular}{ccc}
\begin{minipage}{0.32\linewidth}
\center{\includegraphics[width=\textwidth]{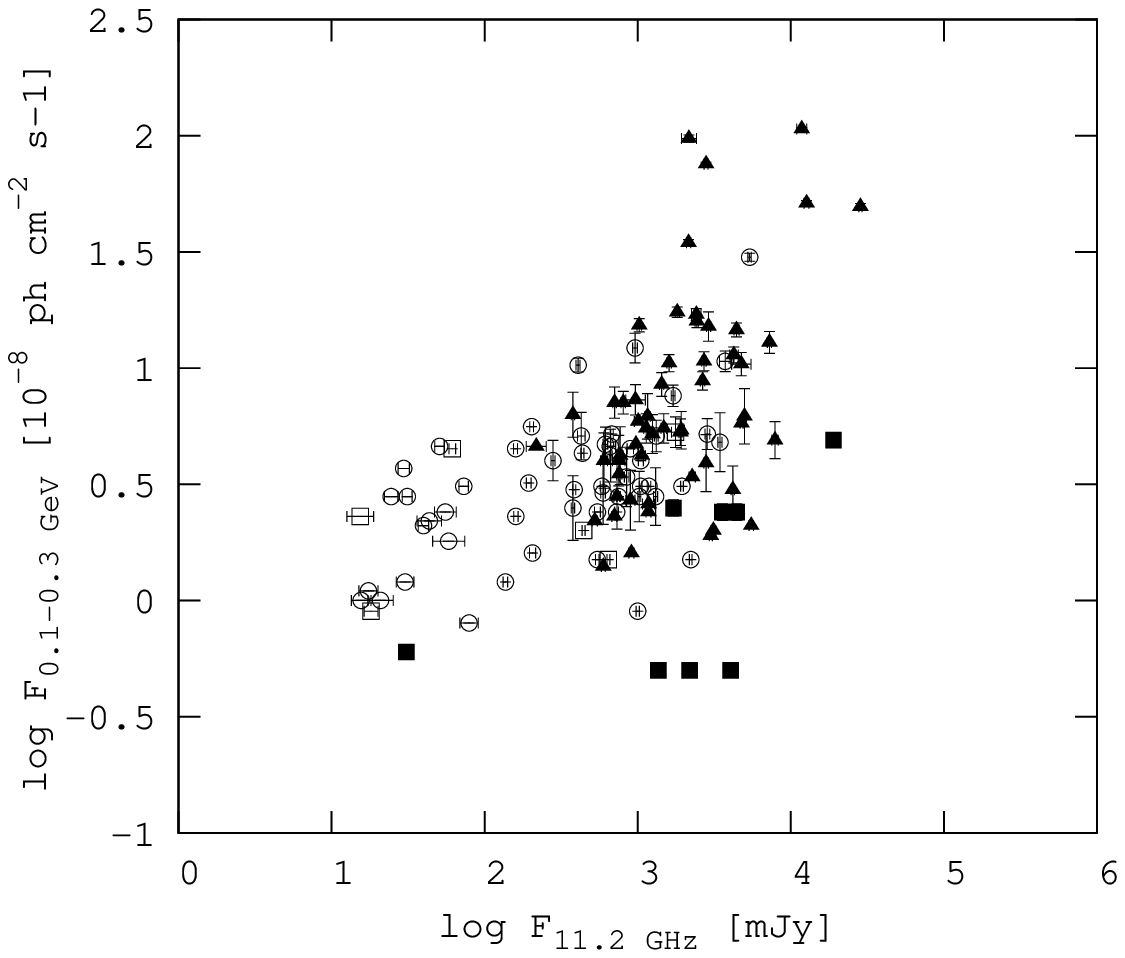}} \\ 
\end{minipage}
\hfill
\begin{minipage}{0.32\linewidth}
\center{\includegraphics[width=\textwidth]{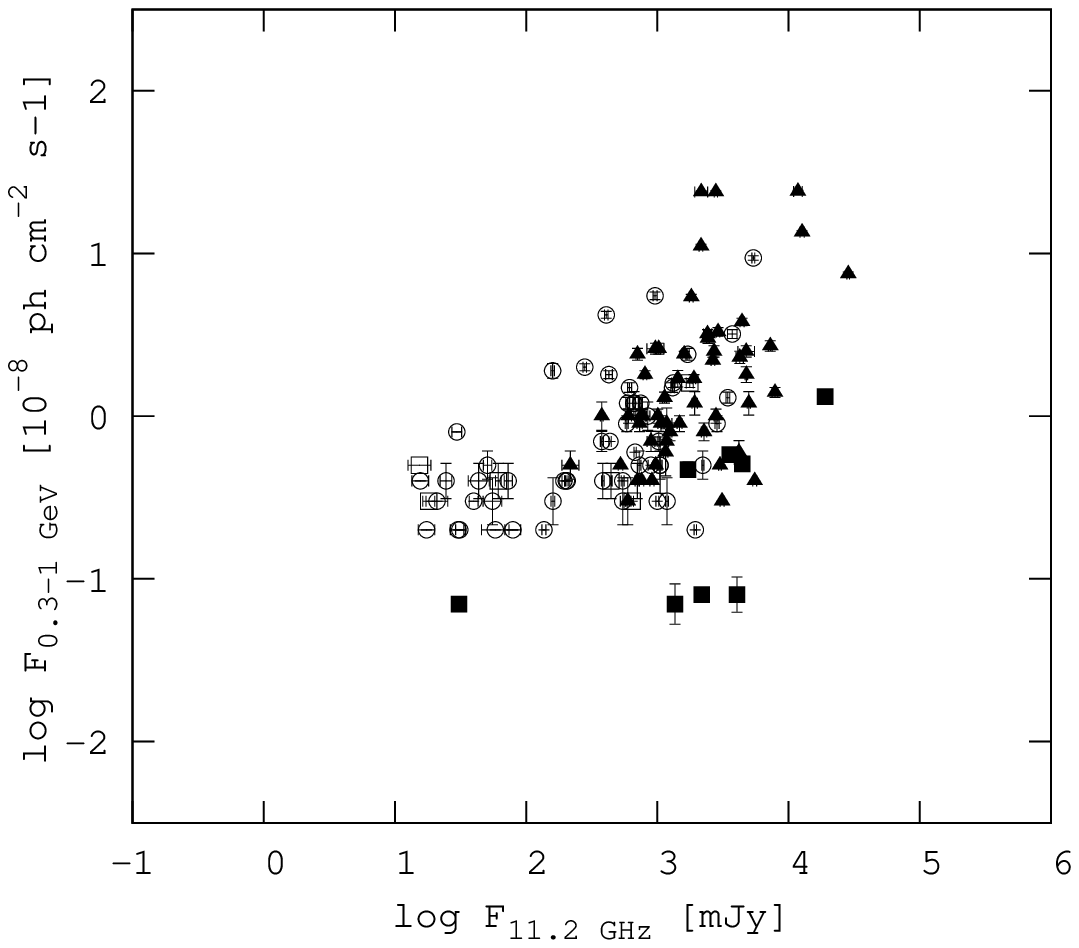}} \\ 
\end{minipage}
\hfill
\begin{minipage}{0.32\linewidth}
\center{\includegraphics[width=\textwidth]{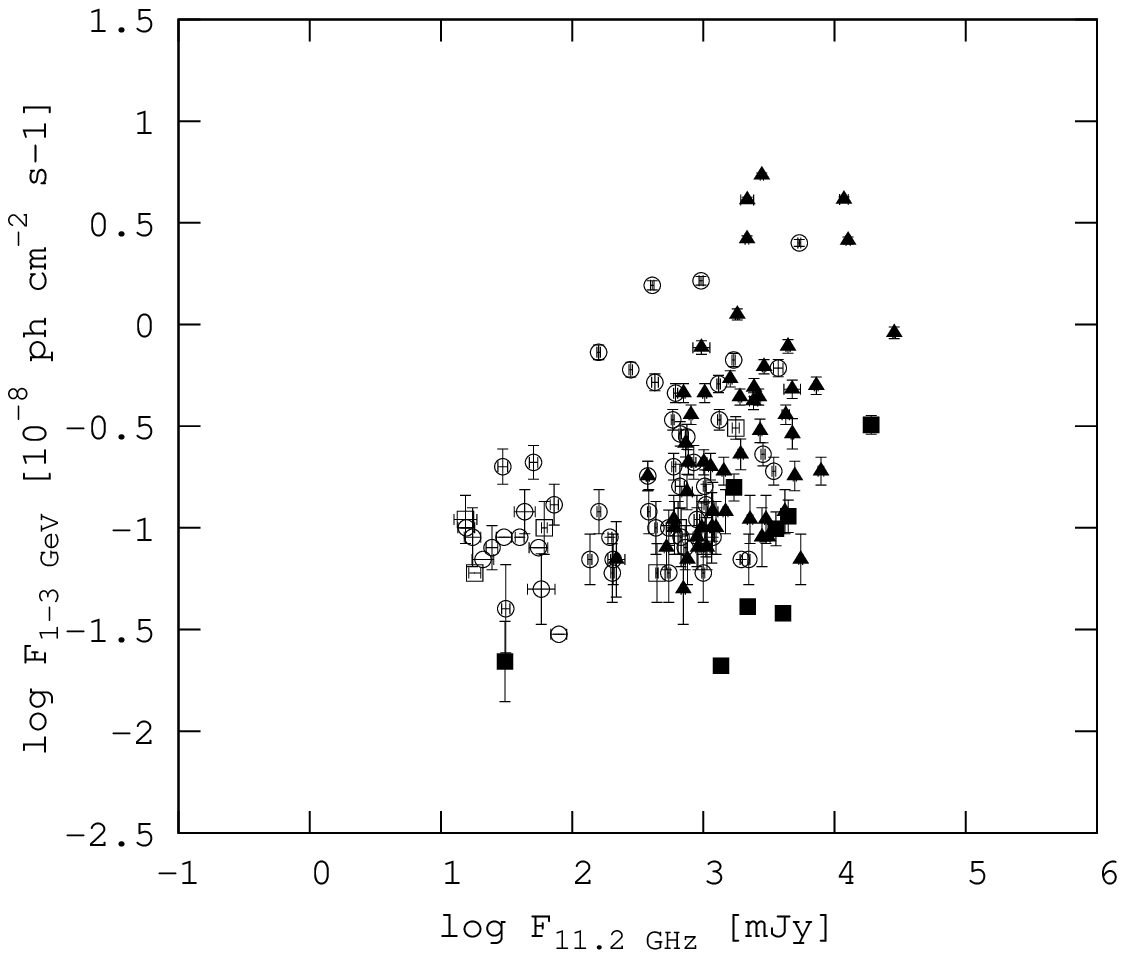}} \\ 
\end{minipage}
\\
\\
\begin{minipage}{0.32\linewidth}
\center{\includegraphics[width=1\linewidth]{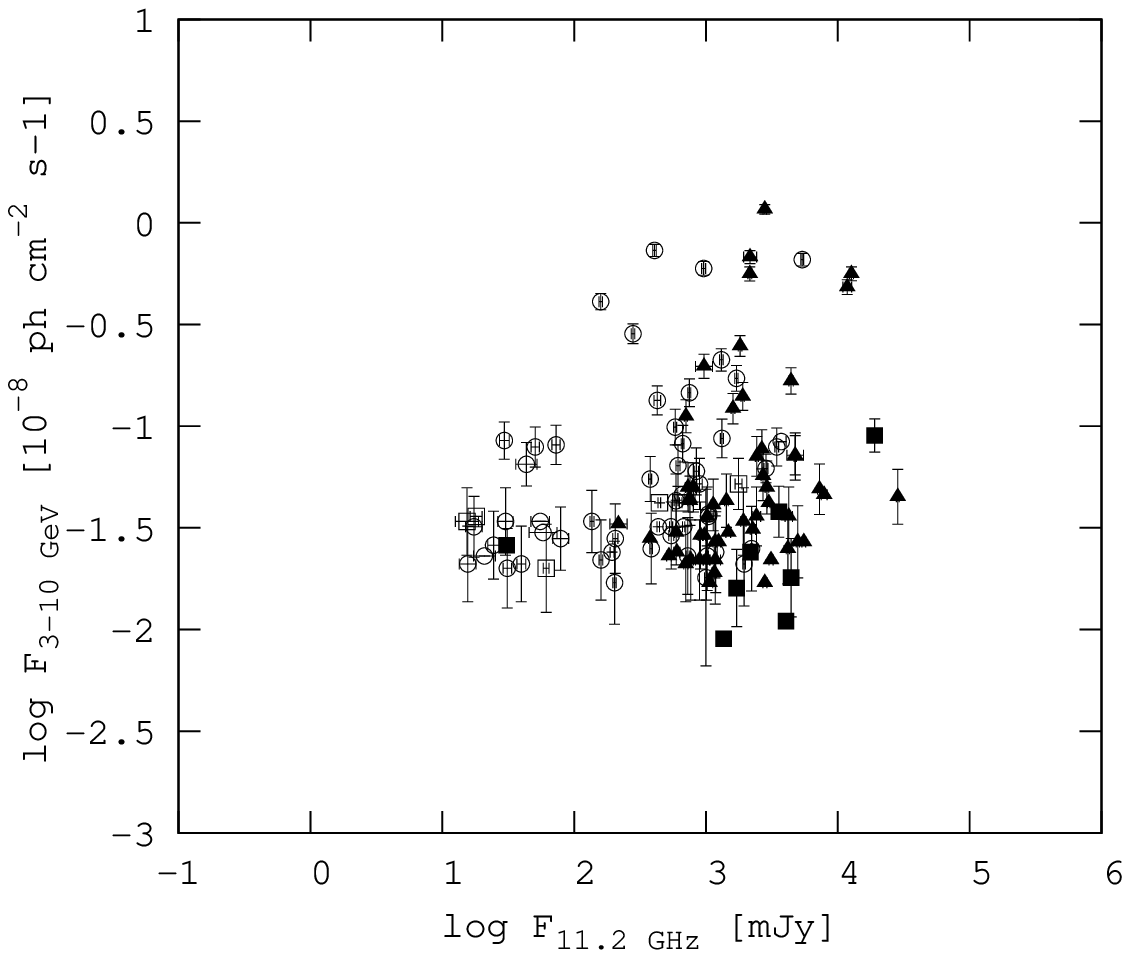}} \\   
\end{minipage}
\begin{minipage}{0.32\linewidth}
\center{\includegraphics[width=1\linewidth]{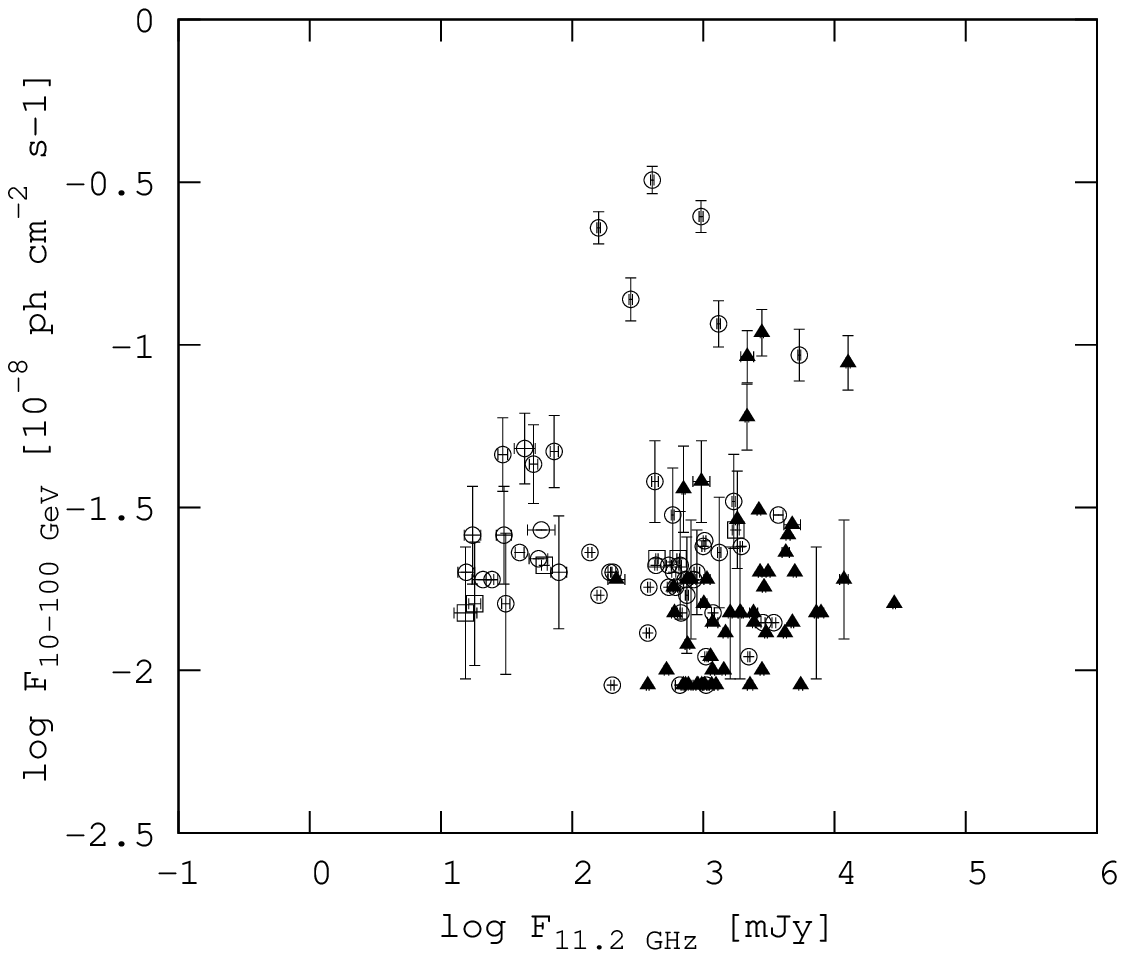}} \\ 
\end{minipage}
\end{tabular}
\caption{Broadband gamma-ray photon flux versus 11.2 GHz flux density. Designations are the same as in Fig \ref{fig1}}
\label{fig2}
\end{figure*}

\begin{figure*}
\centering
\begin{tabular}{ccc}
\begin{minipage}{0.32\linewidth}
\center{\includegraphics[width=\textwidth]{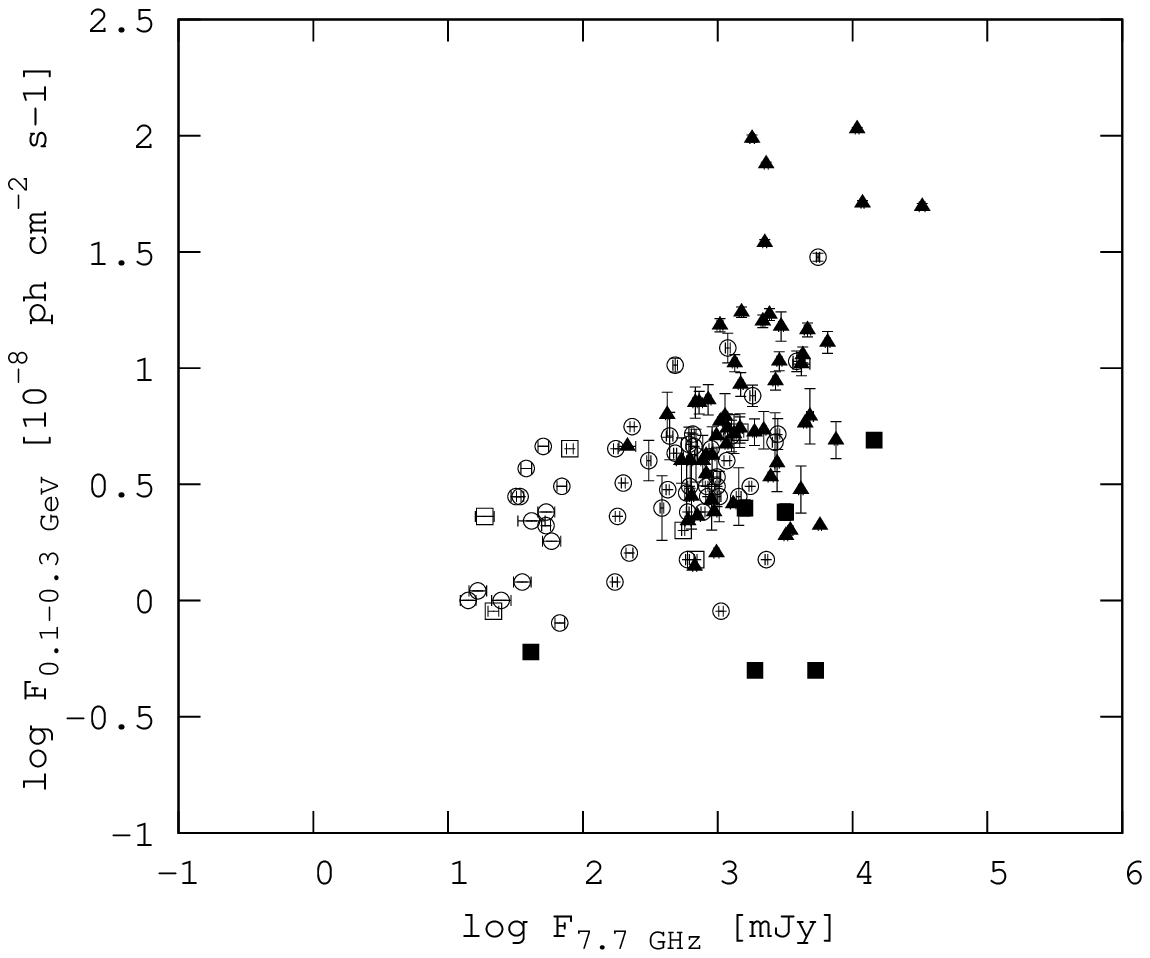}} \\ 
\end{minipage}
\hfill
\begin{minipage}{0.32\linewidth}
\center{\includegraphics[width=\textwidth]{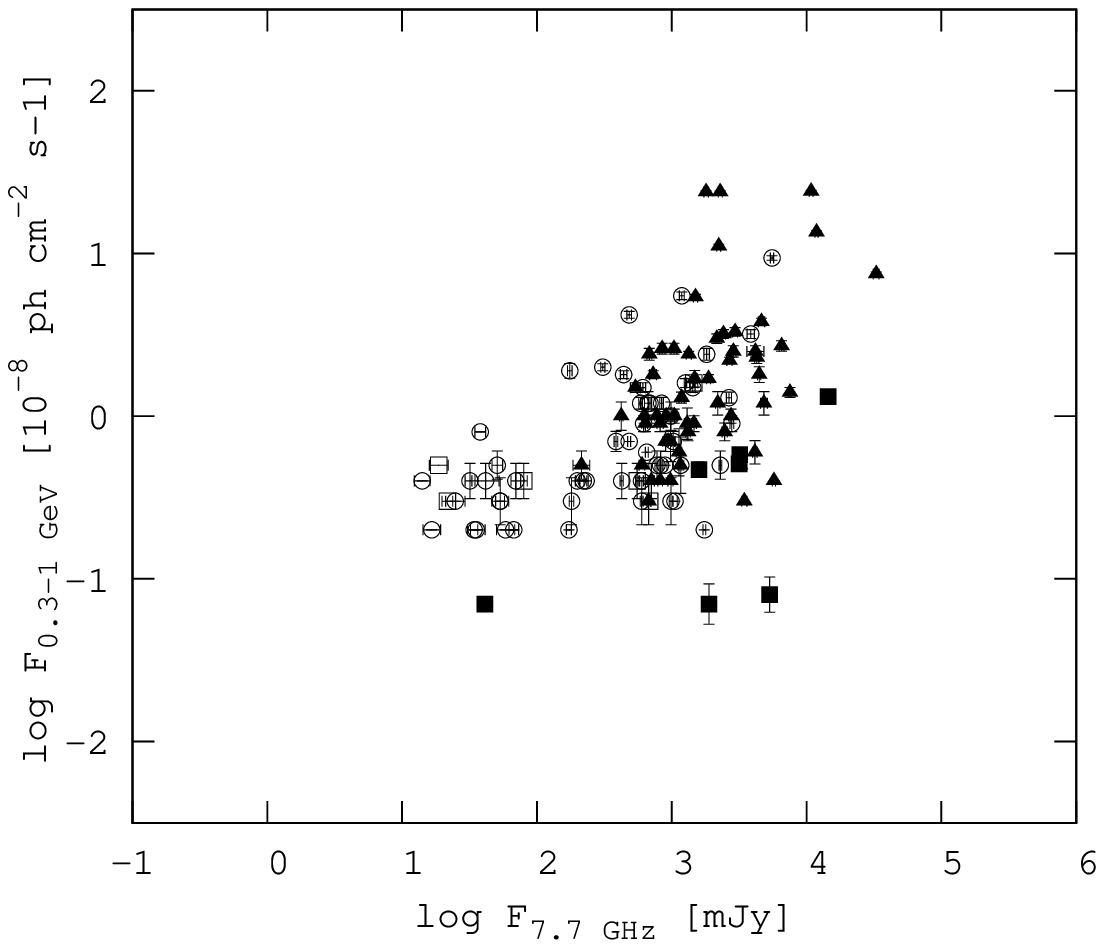}} \\ 
\end{minipage}
\hfill
\begin{minipage}{0.32\linewidth}
\center{\includegraphics[width=\textwidth]{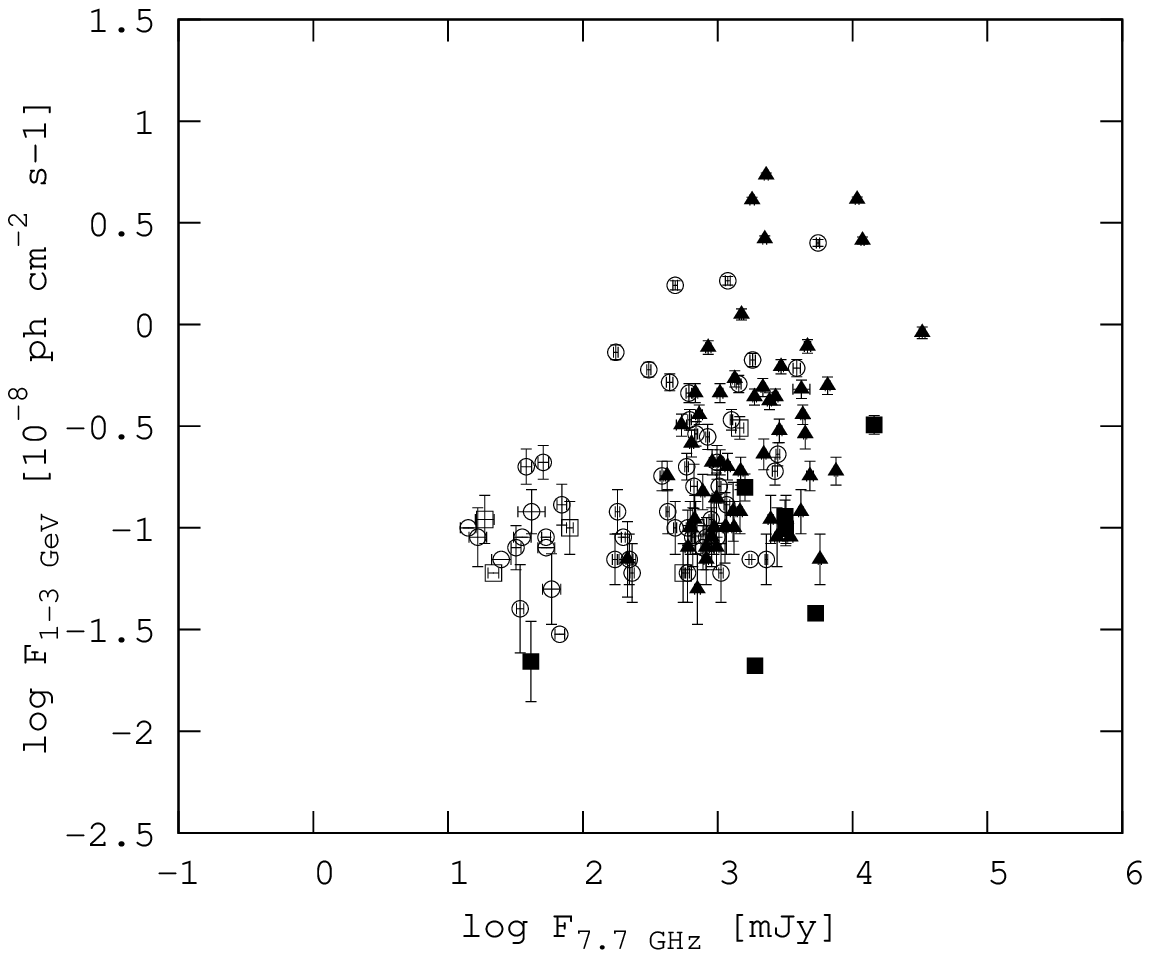}} \\ 
\end{minipage}
\\
\\
\begin{minipage}{0.32\linewidth}
\center{\includegraphics[width=1\linewidth]{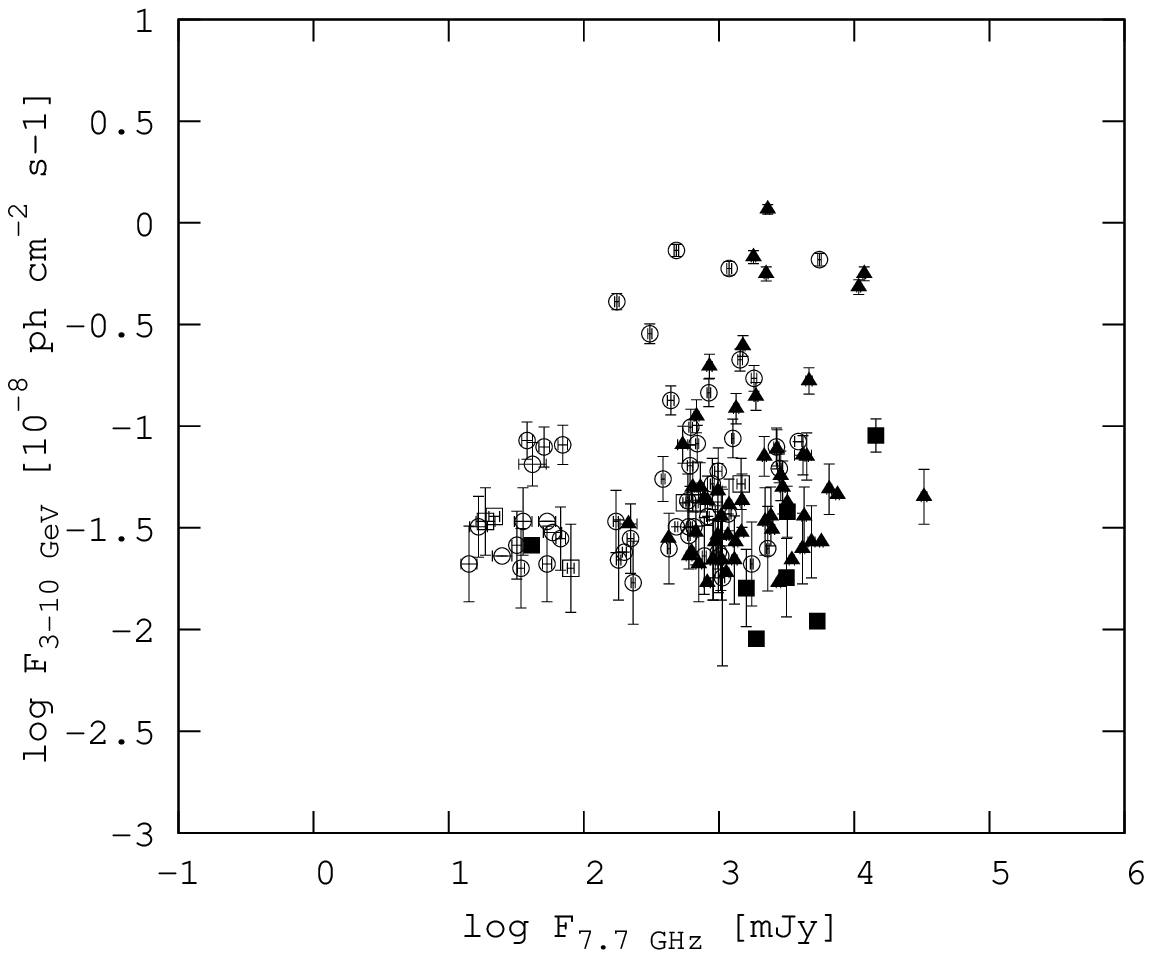}} \\   
\end{minipage}
\begin{minipage}{0.32\linewidth}
\center{\includegraphics[width=1\linewidth]{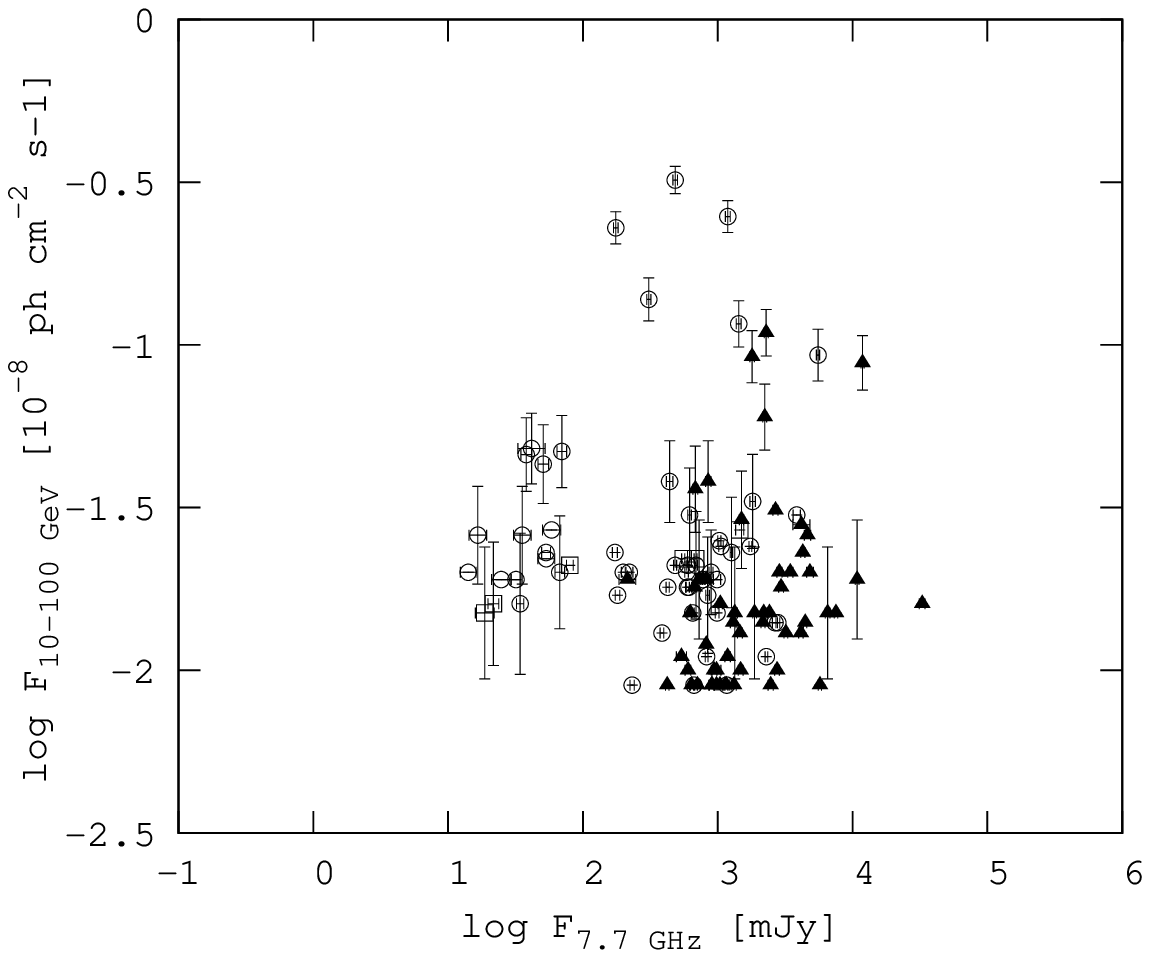}} \\ 
\end{minipage}
\end{tabular}
\caption{Broadband gamma-ray photon flux versus 7.7 GHz flux density. Designations are the same as in Fig \ref{fig1}}
\label{fig3}
\end{figure*}

\begin{figure*}
\centering
\begin{tabular}{ccc}
\begin{minipage}{0.32\linewidth}
\center{\includegraphics[width=\textwidth]{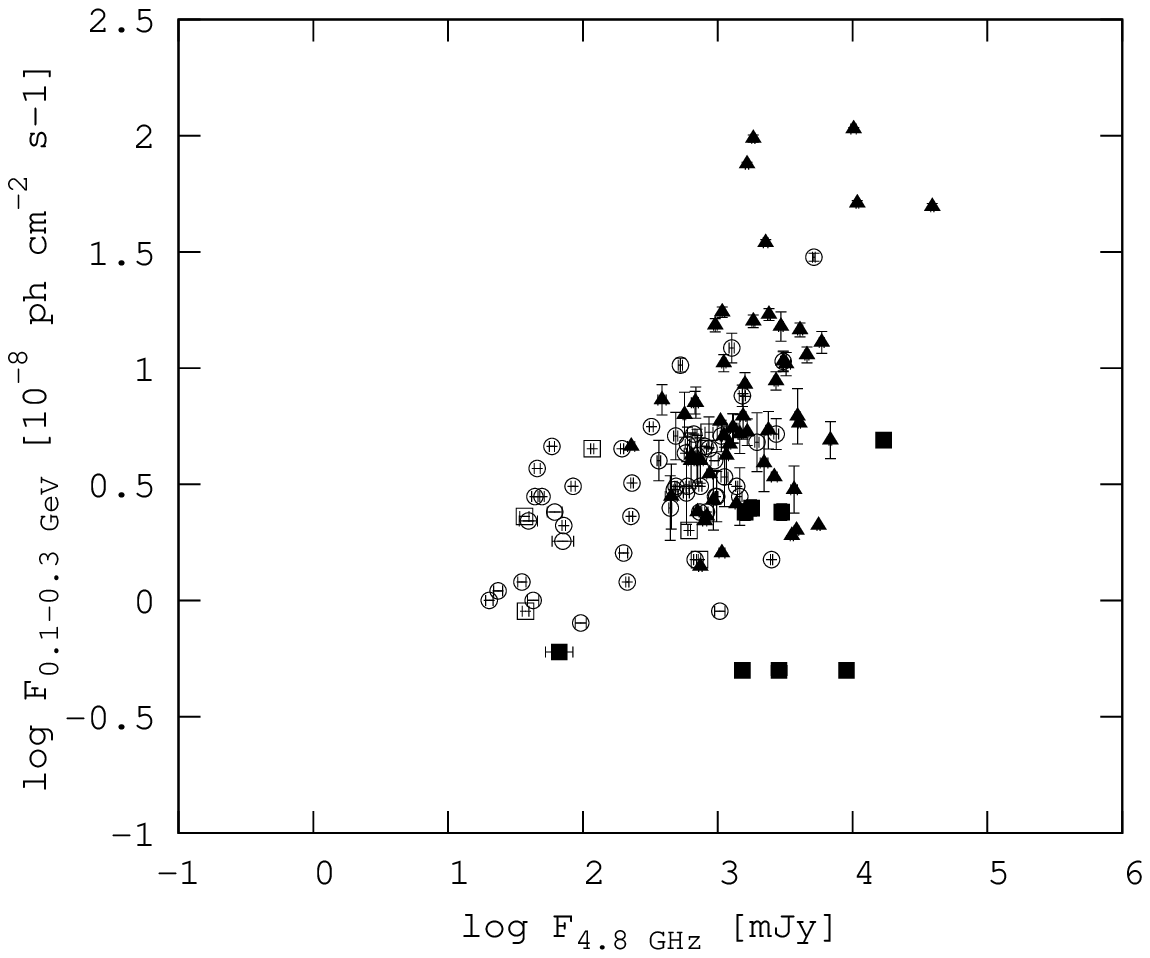}} \\ 
\end{minipage}
\hfill
\begin{minipage}{0.32\linewidth}
\center{\includegraphics[width=\textwidth]{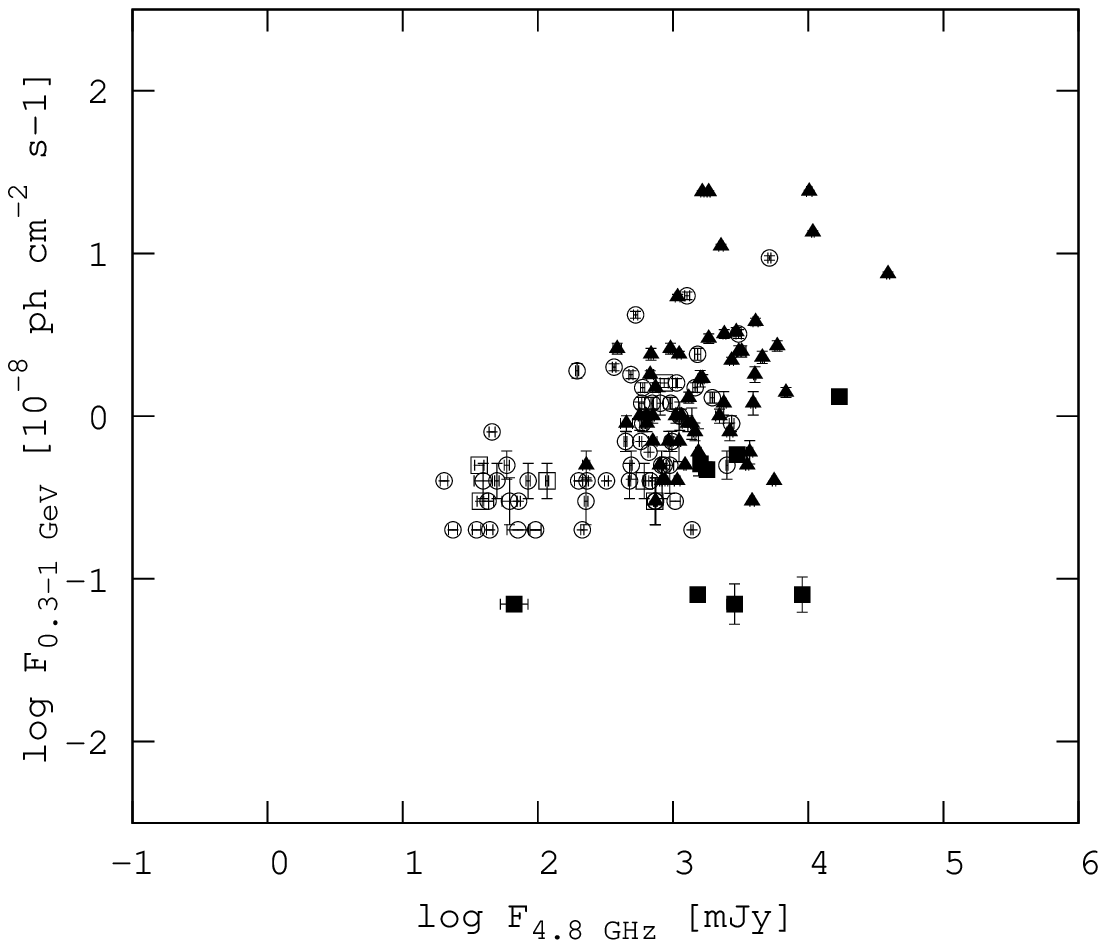}} \\ 
\end{minipage}
\hfill
\begin{minipage}{0.32\linewidth}
\center{\includegraphics[width=\textwidth]{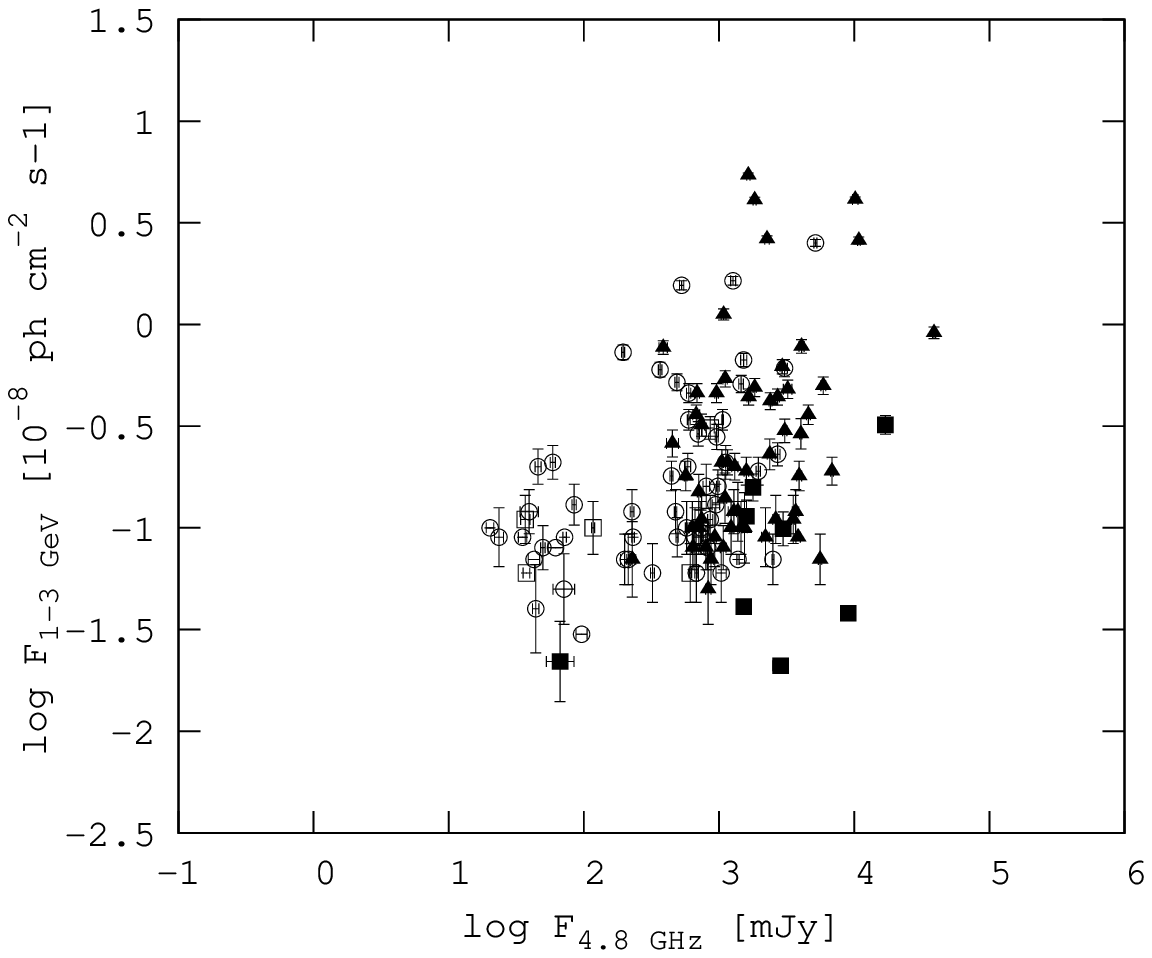}} \\ 
\end{minipage}
\\
\\
\begin{minipage}{0.32\linewidth}
\center{\includegraphics[width=1\linewidth]{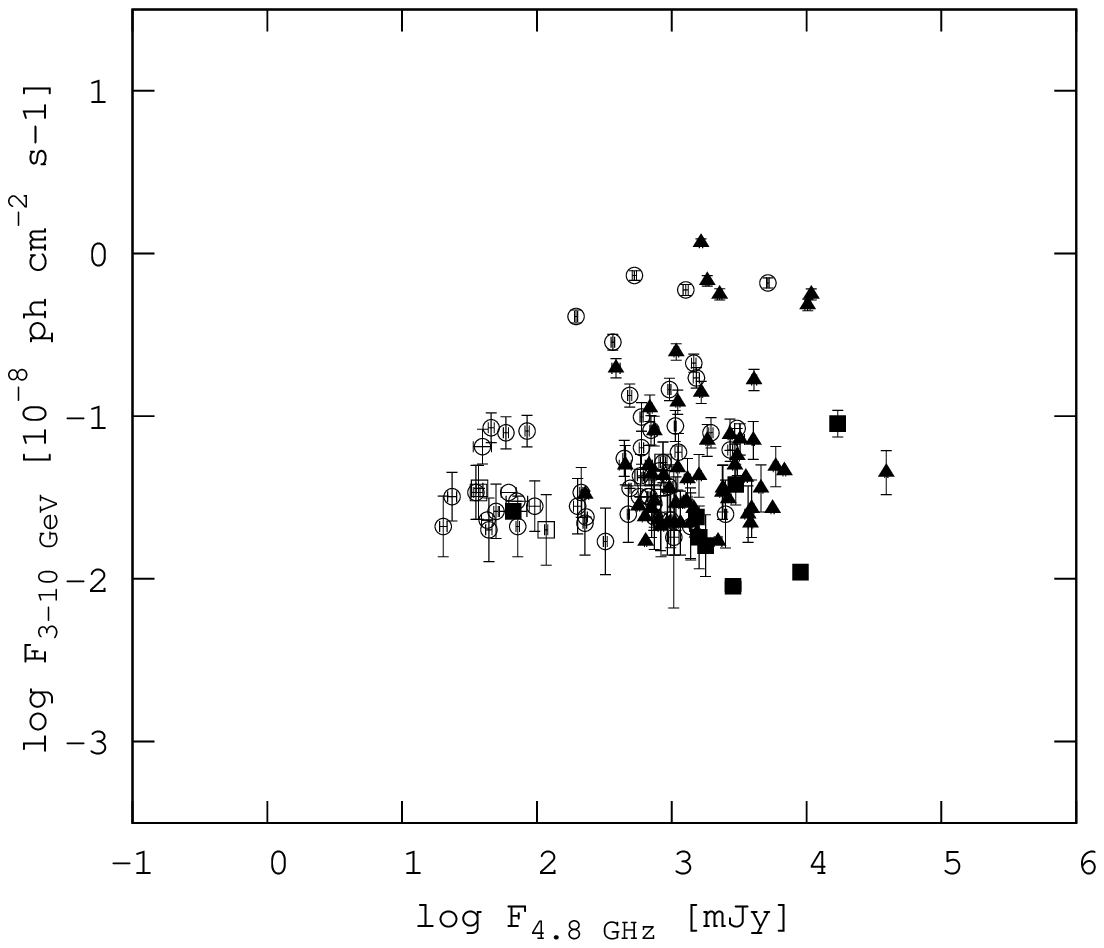}} \\   
\end{minipage}
\begin{minipage}{0.32\linewidth}
\center{\includegraphics[width=1\linewidth]{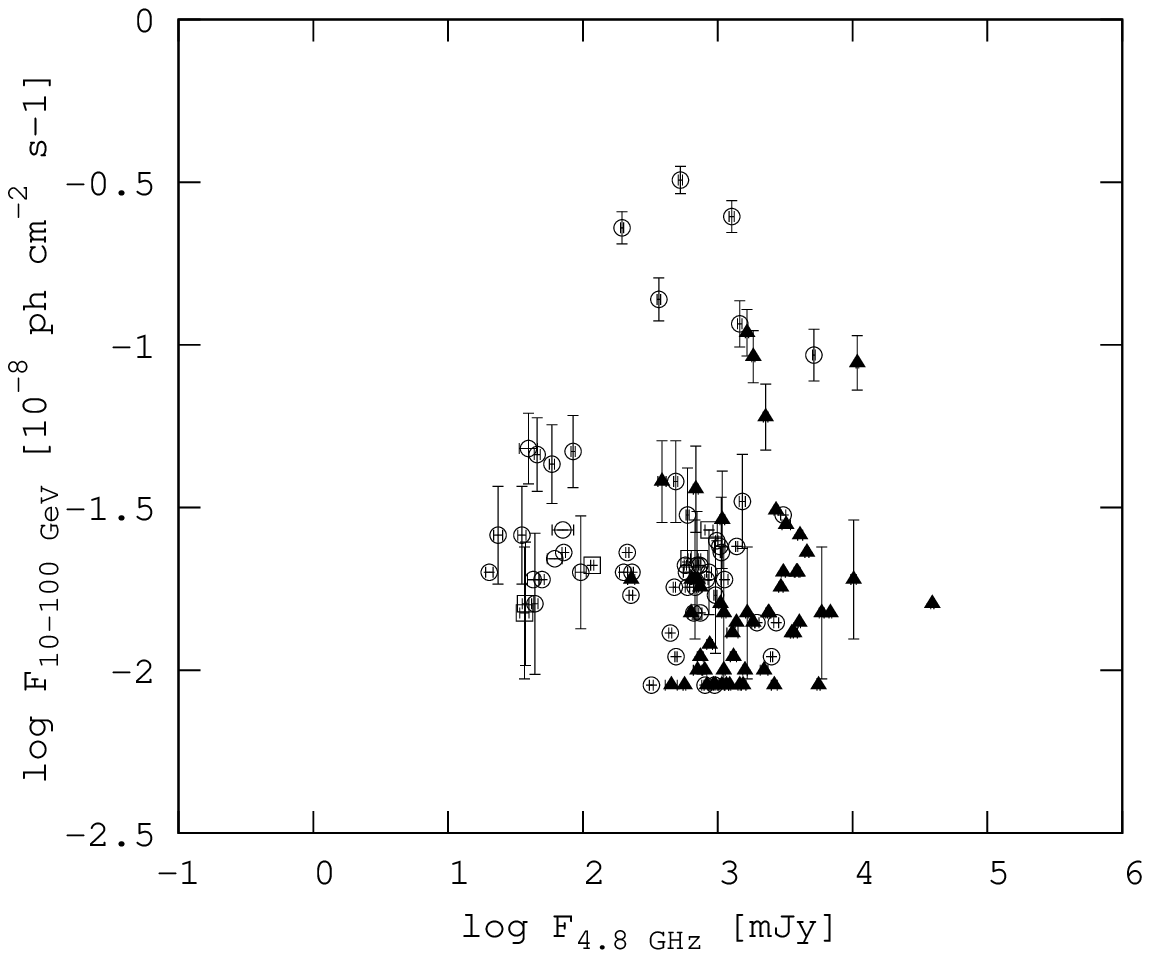}} \\ 
\end{minipage}
\end{tabular}
\caption{Broadband gamma-ray photon flux versus 4.8 GHz flux density. Designations are the same as in Fig \ref{fig1}}
\label{fig4}
\end{figure*}

\begin{figure*}
\centering
\begin{tabular}{ccc}
\begin{minipage}{0.32\linewidth}
\center{\includegraphics[width=\textwidth]{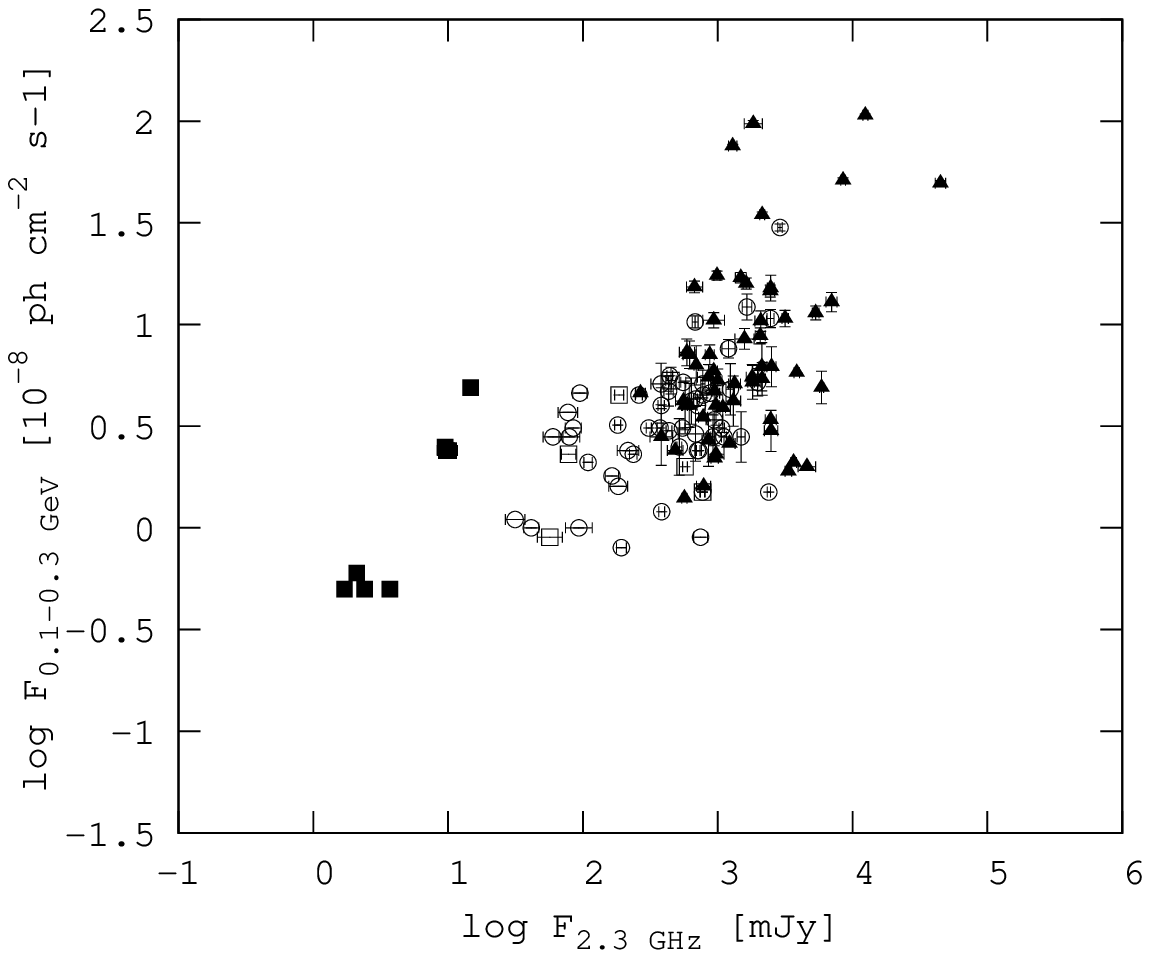}} \\ 
\end{minipage}
\hfill
\begin{minipage}{0.32\linewidth}
\center{\includegraphics[width=\textwidth]{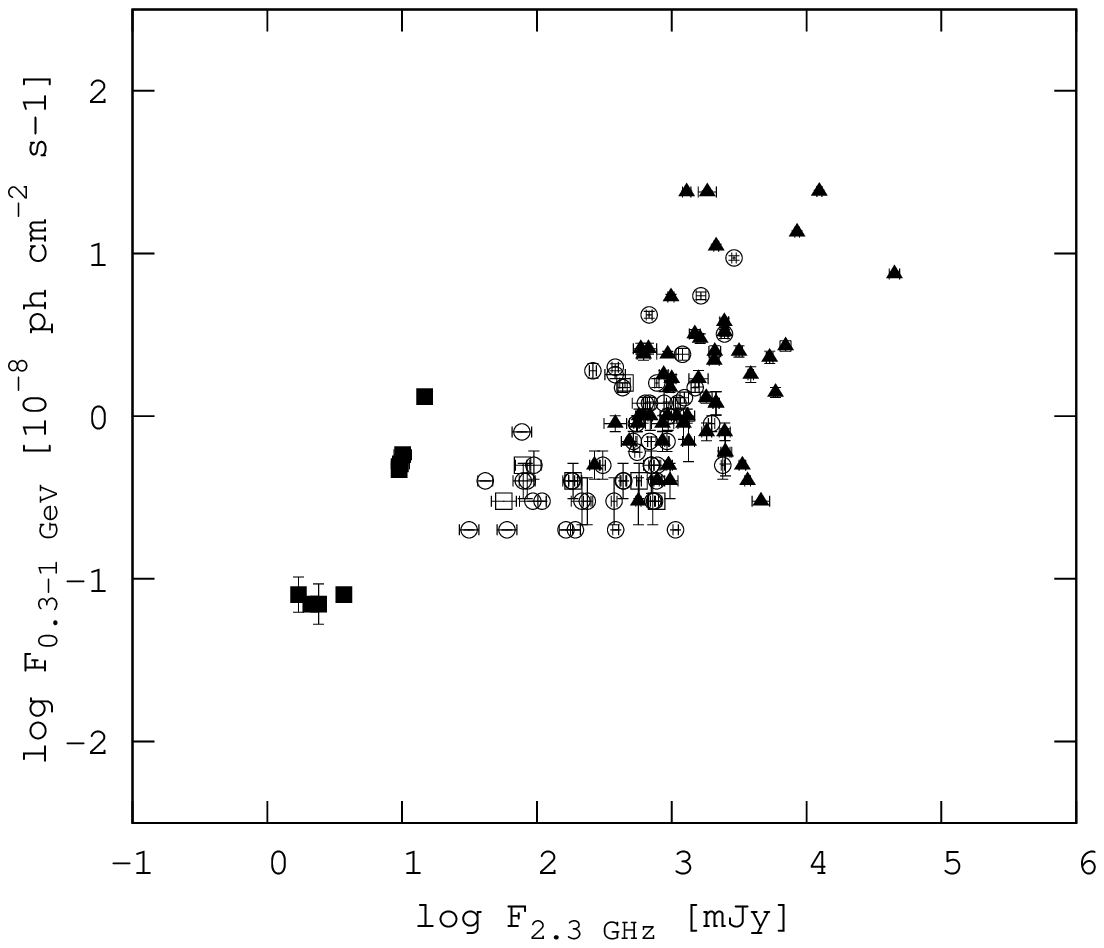}} \\ 
\end{minipage}
\hfill
\begin{minipage}{0.32\linewidth}
\center{\includegraphics[width=\textwidth]{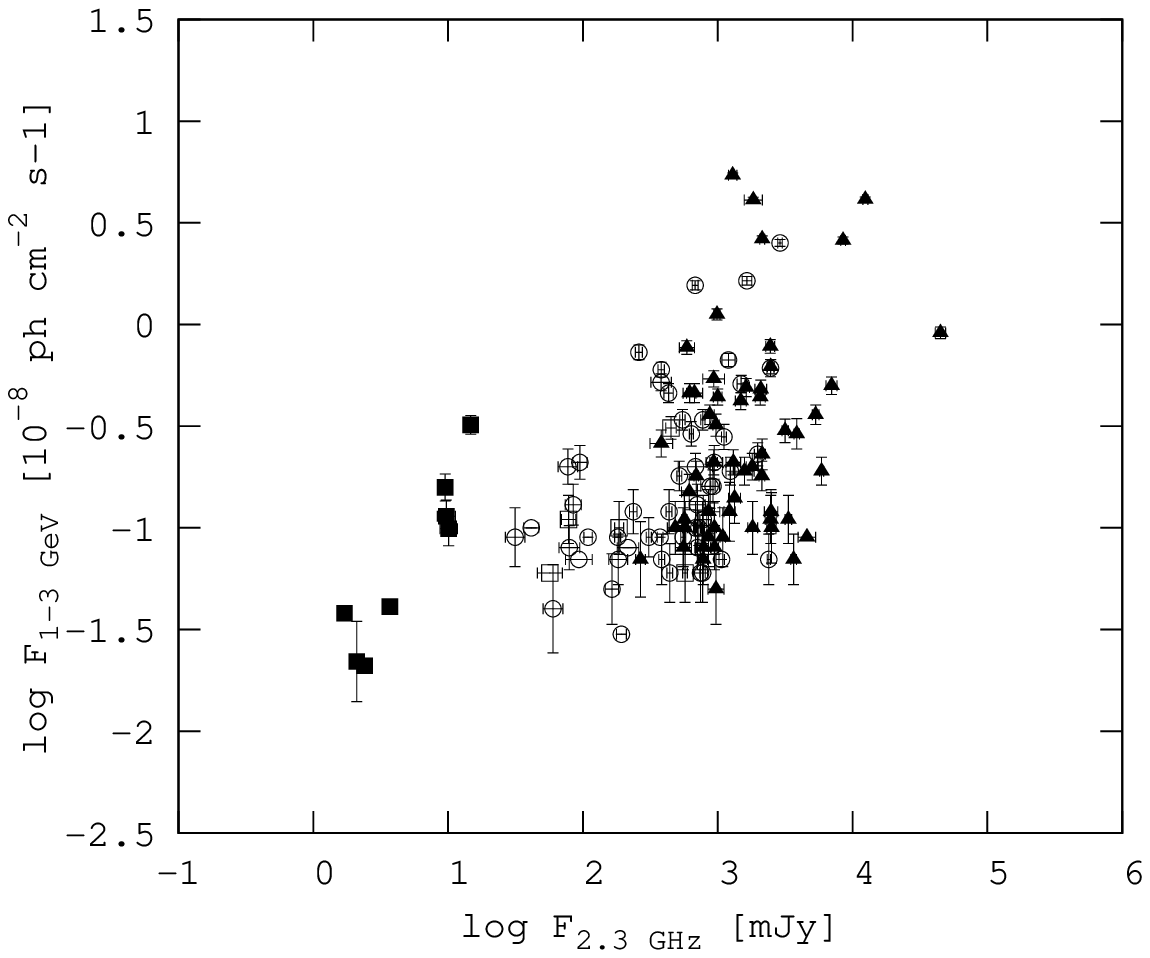}} \\ 
\end{minipage}
\\
\\
\begin{minipage}{0.32\linewidth}
\center{\includegraphics[width=1\linewidth]{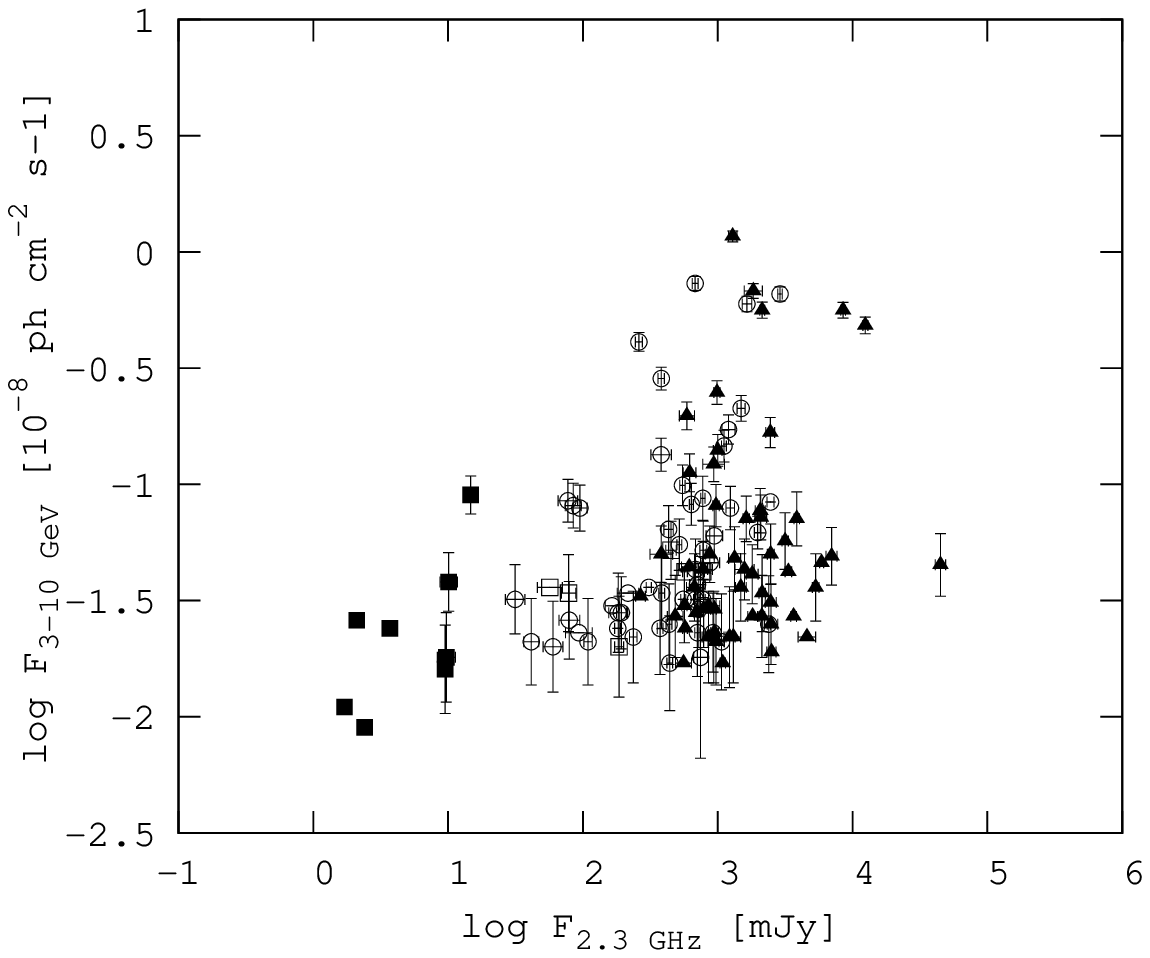}} \\   
\end{minipage}
\begin{minipage}{0.32\linewidth}
\center{\includegraphics[width=1\linewidth]{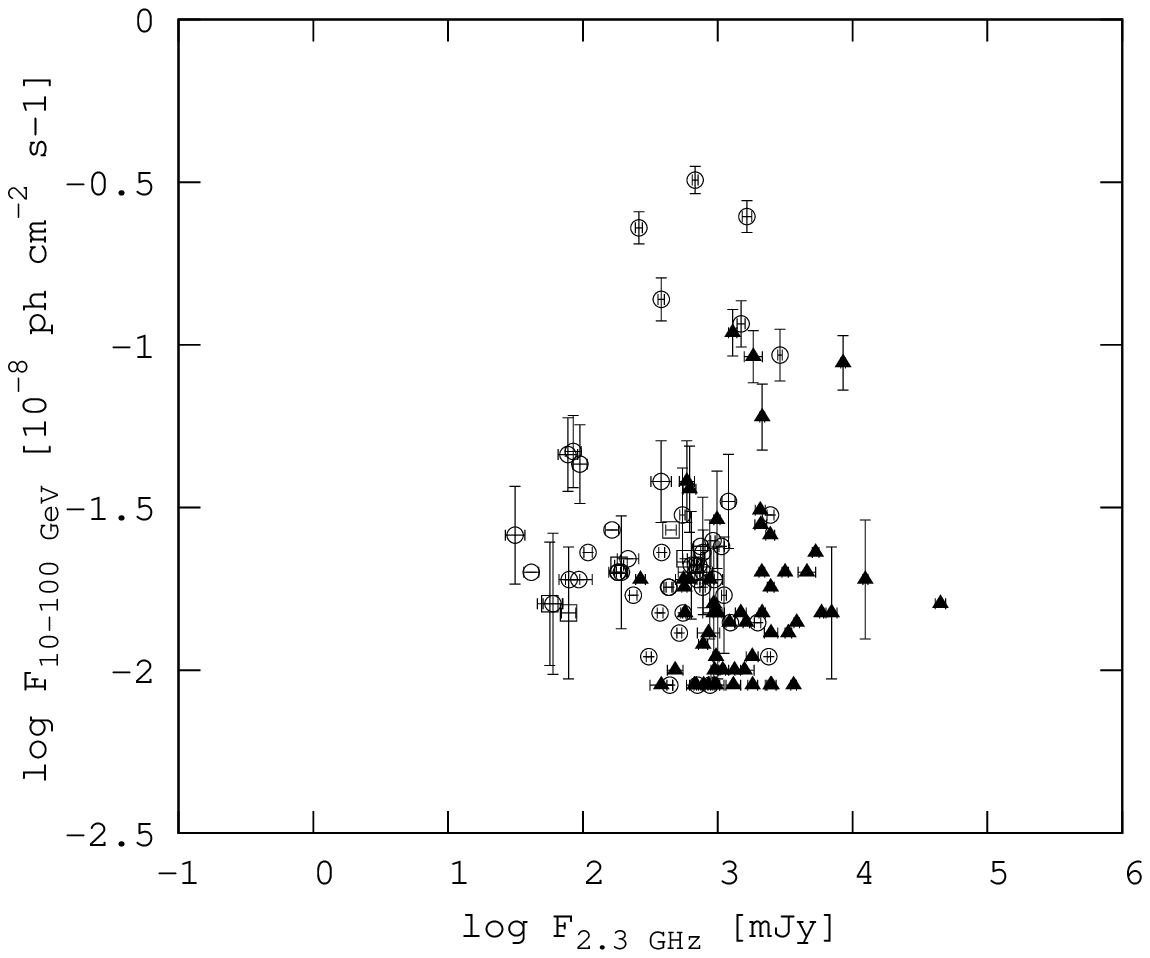}} \\ 
\end{minipage}
\end{tabular}
\caption{Broadband gamma-ray photon flux versus 2.3 GHz flux density. Designations are the same as in Fig \ref{fig1}}
\label{fig5}
\end{figure*}

In Figure \ref{Cor-Energy} we visualize the correlation coefficients, reported in Table \ref{table_corr}.
In this figure, the correlation coefficients are shown across the five energy bands and for each of the five frequencies.
The accuracy, to which the correlation coefficients are determined, is shown by the error bars,
which correspond to the standard deviation defined as in \citet{2003psa..book.....W}:
$\sigma_r = (1-r^2)/(\sqrt{N-1})$
where $N$ is the number of objects and $r$ is the Pearson product-moment correlation coefficient.

The correlation coefficient is sensitive to both radio frequency and gamma-ray energy band.
The correlation coefficient decreases with increasing energy band for both BL Lacs and FSRQs.
Also it can be seen from Figure \ref{Cor-Energy} that this fall is greater for BL Lacs.
On average, correlation coefficients are higher for BL Lacs; and comparable values for BL Lacs and FSRQs are obtained only when the 21.7 GHz frequency band is considered.

For both types of sources we found a common trend: the highest correlation of the flux densities at all radio frequencies was detected with the $0.1-0.3$ GeV gamma-ray band flux. Larger correlation coefficients occur for BL Lacs than for FSRQs at all radio bands (except 21.7 GHz) with the $0.1-0.3$ and $0.3-1$ GeV bands.
At $10-100$ GeV no correlation with any radio band was found for both types of blazars.
Moreover, for FSRQs the correlation coefficient is more stable across the various energy bands
in comparison with BL Lacs, for which Pearson's $r$ decreases faster for higher energy gamma-ray bands.
If we examine correlation strength evolution across radio frequencies, BL Lacs showed a correlation of the same order at both 2.3 and 21.7 GHz with the flux in the $0.1-1$ GeV range.
FSRQs showed noticeably less correlation at 2.3 GHz, compared the correlation at 21.7 GHz, with the flux from the same gamma-ray range.
Hence, we find that for BL Lac blazars, the correlation of the fluxes appears to be more sensitive to the considered gamma-ray energy band than to the frequency, while for FSRQ sources, the correlation changed both with the considered radio frequency and gamma-ray energy band.

\subsection{Robustness test for the correlation}
We used the permutation method described by \citet{2012ApJ...751..149P}
to test the significance of the computed correlation coefficients.
It is intended to take account of common distance bias and Malmquist bias.
Simulated samples contained only sources with known redshifts (because luminosities are used in this method)
and thus the number of objects was reduced to 103 (48 BL Lacs and 55 FSRQs).
We ran the permutation procedure $10^7$ times.
Thus, the probability to obtain Pearson's $r$ value more or equal to the real one
from intrinsically uncorrelated measurements was calculated
 (hereinafter referred to as significance or $sig$). 
The results of the data randomization analysis are presented in Table \ref{table_sig},
where we give the number of sources for each blazar subset (N),
the value of the Pearson's $r$,
and the statistical significance of the apparent correlation ($sig$).
Correlation coefficients are slightly different in some cases,
from those given in Table \ref{table_corr}, because  
 only sources with known redshifts were involved in this robustness test.

We find that the highest significances occurs for higher correlation coefficients (e.g., $r$=$0.75$ with the chance probability $sig$=$3.24\times 10^{-7}$).
The least significant cases are observed for the $10-100$ GeV band for both types of blazar.
Also there is a high probability for the correlation
 appearing by chance in the $1-3$ and $3-10$ GeV bands for FSRQs (up to 56 \%).
BL Lacs demonstrate only marginally significant correlation
between the flux densities at 11.2 and 21.7 GHz and the photon flux at the $3-10$ GeV band 
(2-6 \% probability of achieving same or higher correlation from intrinsically uncorrelated data).

\section{Discussion and Conclusions}
We searched for a possible correlation between gamma-ray and radio emission in blazars. 
We used quasi-simultaneous multi-frequency \textit{Fermi}-LAT and RATAN-600 observational data to test for the existence of the connection between fluxes in these bands.
The correlation analysis was performed using the flux densities from five radio bands ($2.3 - 21.7$ GHz) for the first time.
If the gamma-ray emission is produced by the 
SSC method, we would expect a tight correlation
between gamma-ray and radio emission because the seed photons are produced by the same method.
In previous work it is shown that SSC models are acceptable for most BL Lacs SEDs,
 while FSRQs usually require an external inverse Compton mechanism (e.g., \citealt{2005A&A...440..845L,2010ApJ...716...30A,2014MNRAS.439..690H}).
Since SSC would imply the same origin for gamma-ray and radio emission, generally higher values of the correlation coefficient for BL Lacs were expected.

We found a positive apparent correlation between gamma-ray photon flux and radio flux density for sources from our sample,
but with considerable scatter (Pearson's $r$ ranges from -0.03 to 0.74 for BL Lacs and from 0.03 to 0.57 for FSRQs).
The BL Lacs show higher values of the correlation coefficient than FSRQs at all frequencies, except 21.7 GHz, and at all bands, except $10-100$ GeV, when comparing different sub-samples of blazars.
Also BL Lacs showed the correlation of the same order at both 2.3 and 21.7 GHz with the flux from the $0.1-1$ GeV range, 
while FSRQs demonstrated noticeably less correlation at 2.3 GHz, compared with the correlation at 21.7 GHz, with the flux from the same gamma-ray range.
At higher gamma-ray energies, the correlation weakens and even becomes negative for both BL Lacs and FSRQs,
consistent with \citet{2011ApJ...741...30A}.
The strength of the correlation clearly depends on the gamma-ray energy band considered for both types of blazars; the correlation of the radio flux density at any frequency is stronger with the $0.1-1$ GeV energy range.
Thus, for BL Lac blazars, the correlation of the fluxes appeared to be more sensitive to the considered gamma-ray energy band, than to the frequency, while for FSRQ sources correlation seems to change both with the considered radio frequency and gamma-ray energy band.

We used data randomization tests to quantify the significance of the computed correlation coefficients.
We find that the statistical significance of correlations we obtained between the flux densities at all frequencies 
and the photon flux at all gamma-ray bands below 3 GeV is high for BL Lacs (chance probability $\sim 10^{-3} - 10^{-7}$).
The correlation coefficient is highly significant for $0.1-0.3$ GeV and low and insignificant for the $10-100$ GeV band
for both types of blazars for all considered frequencies.

We used quasi-simultaneous data at radio frequencies; there was 
RATAN-600 data for the first year of operation of \textit{Fermi}-LAT (within a few months).
Simultaneity of the data used plays important role in analysis and as shown in previous works (e.g., \citealt{2011ApJ...741...30A,2012ApJ...754...23L}) strength of the correlation depends on concurrency.
\citet{2011ApJ...741...30A} reveal stronger correlation between gamma-ray and radio emission for concurrent data than for archival (comparing archival 8.4 GHz data and simultaneous 15 GHz measurements).
\citet{2012ApJ...754...23L} found higher statistical significance for the $L_{\gamma}$ - $L_{radio}$ correlation in data obtained quasi-simultaneously (within months), than in simultaneous observations (within weeks) and in data averaged over a long-term period (27 months).
\citet{2014MNRAS.441.1899F} analysed gamma-ray/radio light curves of
54 \textit{Fermi}-bright blazars. They estimated $12\pm 8$ d time lag at 3 mm wavelength.
According to their study, the time delay systematically increases from the millimetre to the centimetre band.
If the same beamed relativistic electrons first up-scatter
the synchrotron photons to gamma-ray energies
and then flare in the radio band, then a significant time delay
between the radio and gamma-ray flares is expected.
Also such time lags would occur, if 
a disturbance (a shock, propagating along the jet) appears from the radio core
and first causes a flare in the (sub-)millimetre band and
 then produces a gamma-ray flare downstream in the jet.
However, there is no commonly accepted physical model for the radio/gamma-ray emission regions:
some studies suggest gamma-ray flare occurring downstream from the radio core,
others, upstream from the core.
Scenarios vary depending on the flare type (relatively strong or not),
source type (BL Lacs and quasars) and radio band (mm, cm).
The possible locations of the gamma-ray emission site are discussed in, for example, \citet{2011A&A...532A.146L} and \citet{2014MNRAS.441.1899F} and references therein.
Nonetheless, it is well known that there is a considerable delay 
between flares in radio and gamma-ray bands.
That is why there are might be a larger chance of detecting a correlation 
in averaged radio and gamma-ray fluxes, than in simultaneous observational data. 

Together with the previous results of the correlation between gamma-ray and radio band emission for AGNs \citep{2010MNRAS.407..791G,2009ApJ...696L..17K,2011ApJ...741...30A}, our results could be considered as suggesting the existence of a relationship between the emission from these bands, and as indication of the common origin (in terms of the assumed SSC radiation).
Two possible radiation mechanisms, responsible for gamma-ray emission and SSC and external inverse Compton processes, probably always occur together, although in different proportions.
Hence, it is difficult to provide sufficient reasons to conclude there is a direct and obvious link between gamma-ray and radio band emission in blazars.

\section*{Acknowledgements}
The RATAN-600 observations were carried out with the
financial support of the Ministry of Education and Science of the Russian
Federation (state contract no.14.518.11.7054) and Russian Foundation for Basic Research (project no. 12-02-31649).

We thank the anonymous referee for useful comments and suggestions.
The authors want to acknowledge S. Trushkin, V. Stolyarov and C. Riseley for the careful reading and help in preparing this manuscript.
M. Mingaliev acknowledges support through the Russian Government Program of Competitive Growth of the Kazan Federal University.

We used on-line version of the Roma-BZCAT catalogue at the ASI Science Data Center (ASDC) website;
therefore we are grateful to the ASDC staff.

This research made use of the NASA/IPAC Extragalactic Database (NED), which is operated by the Jet Propulsion Laboratory, California Institute of Technology, under contract with the National Aeronautics and Space Administration.

This research also has made use of the Vizier catalogue access tool, CDS, Strasbourg, France.

\bibliographystyle{mn2e-long} 
\bibliography{biblioteka} 

\begin{center}

\begin{table*}
  \caption{List of objects}
  \label{table}
  \begin{tabular}{lcclcc}
    \hline
    Name & Type  & z & Name & Type  & z \\
    \hline
    
    BZB J0022+0608 & BL Lac & -     & PKS 1118-05 & FSRQ  & 1.297 \\
    BZB J0035+1515 & BL Lac & 1.28  & FBQS J115019.2+241753 & BL Lac & 0.2 \\
    GC0039+23 & FSRQ  & 1.426 & 4C 29.45 & FSRQ  & 0.725 \\
    PKS 0047+023 & BL Lac & 1.44  & EXO1218.8+3027 & BL Lac & 0.182 \\
    FBQS J0050-0929 & BL Lac & 0.103 & ON 231 & BL Lac & 0.102 \\
    PKS 0106+01 & FSRQ  & 2.099 & PKS 1219+04 & FSRQ  & 0.966 \\
    GC 0109+224 & BL Lac & 0.265 & PKS 1222+21 & FSRQ  & 0.435 \\
    BZQ J0136+4751 & FSRQ  & 0.859 & 3C 273 & FSRQ  & 0.158 \\
    PKS 0139-09 & BL Lac & 0.733 & FBQS J123014.0+251807 & BL Lac & 0.135 \\
    2MASX J01593439+1047052 & BL Lac cand & 0.195 & 3C 279 & FSRQ  & 0.53 \\
    4C +15.05 & Blaz.uncer & 0.833 & 1WGA J1310.4+3220 & FSRQ  & 0.997 \\
    2MASS J02171711+0837038 & BL Lac cand & 1.4   & GB6 J1327+2210 & FSRQ  & - \\
    PKS 0215+015 & FSRQ  & 1.715 & PKS 1335-127 & FSRQ  & 0.539 \\
    B2 0218+35 & Blaz.uncer & 0.944 & FIRST J134105.1+395945 & BL Lac & 0.172 \\
    3C 66A & BL Lac & 0.444 & CGRaBS J1357+7643 & FSRQ  & 1.585 \\
    BZQ J0237+2848 & FSRQ  & 1.213 & PKS 1406-076 & FSRQ  & 1.494 \\
    PKS 0235+164 & BL Lac & 0.94  & FBQS J142700.4+234800 & BL Lac & 0.16 \\
    2MASX J02503793+1712092 & Blaz.uncer & 1.1   & 2MASS J14424821+1200402 & BL Lac & 0.162 \\
    BZB J0303+4716 & BL Lac & 0.475 & PKS 1502+106 & FSRQ  & 1.839 \\
    BZB J0316+0904 & BL Lac & -     & PKS 1502+036 & FSRQ  & 0.409 \\
    NGC 1275 & Blaz.uncer & 0.018 & PKS 1510-08 & FSRQ  & 0.36 \\
    BZB J0319+1845 & BL Lac & 0.19  & PKS 1514+197 & BL Lac & 0.65 \\
    1H 0323+022 & BL Lac & 0.147 & 1RXS J152239.7-273025 & BL Lac & 1.294 \\
    PKS 0332-403 & BL Lac cand & -     & 4C +05.64 & FSRQ  & 1.422 \\
    NRAO 140 & FSRQ  & 1.259 & BZB J1555+1111 & BL Lac & 0.36 \\
    PKS 0336-017 & FSRQ  & 0.85  & SDSS J160706.23+155136.8 & BL Lac & 0.357 \\
    1H 0413+009 & BL Lac & 0.287 & 4C +10.45 & FSRQ  & 1.226 \\
    PKS 0420+022 & FSRQ  & 2.277 & BZQ J1613+3412 & FSRQ  & 1.399 \\
    PKS 0420-01 & FSRQ  & 0.916 & 4C +38.41 & FSRQ  & 1.813 \\
    PKS 0422+004 & BL Lac & 0.31  & 3C 345 & FSRQ  & 0.593 \\
    PKS 0446+11 & BL Lac & 1.207 & MRK 0501 & BL Lac & 0.033 \\
    PKS 0454-234 & FSRQ  & 1.003 & PKS 1717+177 & BL Lac & 0.137 \\
    4C-02.19 & FSRQ  & 2.291 & 2MASS J17250434+1152155 & BL Lac & 0.018 \\
    BZQ J0505+0459 & BL Lac cand & 0.027 & NRAO 530 & FSRQ  & 0.902 \\
    2MASS J05075617+6737242 & BL Lac & 0.314 & 1ES 1741+196 & BL Lac & 0.083 \\
    BZB J0509+0541 & BL Lac & 0.304 & OT 081 & Blaz.uncer & 0.322 \\
    PKS 0507+17 & FSRQ  & 0.416 & BZB J1756+5522 & BL Lac & 0.407 \\
    PKS 0539-057 & FSRQ  & 0.839 & BZB J1800+7828 & BL Lac & 0.68 \\
    OH-10 & FSRQ  & 0.872 & 3C 380.0 & Blaz.uncer & 0.695 \\
    2MASS J06251826+4440014 & BL Lac & -     & BZQ J1852+4855 & FSRQ  & 1.25 \\
    2MASX J07103005+5908202 & BL Lac & 0.125 & PKS 1954-388 & FSRQ  & 0.63 \\
    PKS 0723-008 & Blaz.uncer & 0.128 & BZB J2005+7752 & BL Lac & 0.342 \\
    PKS 0735+17 & BL Lac & 0.424 & PKS 2012-017 & BL Lac & 0.52 \\
    PKS 0736+01 & FSRQ  & 0.189 & BZQ J2035+1056 & FSRQ  & 0.601 \\
    PKS 0748+126 & FSRQ  & 0.889 & PKS 2047+039 & BL Lac cand & - \\
    PKS 0754+100 & BL Lac & 0.266 & PKS 2131-021 & BL Lac & 0.557 \\
    PKS 0805-07 & FSRQ  & 1.837 & BZQ J2143+1743 & FSRQ  & 0.213 \\
    PKS 0808+019 & BL Lac & 0.93  & 4C 06.69 & FSRQ  & 0.999 \\
    B3 0814+425 & BL Lac & 0.53  & PKS 2149+173 & BL Lac & 0.871 \\
    PKS 0823+033 & BL Lac & 0.506 & BL Lac & BL Lac & 0.069 \\
    BZQ J0830+2410 & FSRQ  & 0.941 & PKS 2201+171 & FSRQ  & 1.076 \\
    PKS 0829+046 & BL Lac & 0.18  & PKS 2209+236 & FSRQ  & 1.125 \\
    2EG J0852-1237 & FSRQ  & 0.566 & 3C 446 & FSRQ  & 1.404 \\
    PKS 0851+202 & BL Lac & 0.306 & BZQ J2229-0832 & FSRQ  & 1.56 \\
    BZQ J0920+4441 & FSRQ  & 2.186 & 4C -11.69 & FSRQ  & 1.037 \\
    2MASS J09303759+4950256 & BL Lac & 0.188 & 3C 454.3 & FSRQ  & 0.859 \\
    OK 290 & FSRQ  & 0.708 & PKS 2254-204 & BL Lac & - \\
    2MASS J10121335+0630569 & BL Lac & 0.727 & PKS2255-282 & FSRQ  & 0.926 \\
    SDSS J101603.13+051302.3 & FSRQ  & 1.713 & BZB J2304+3705 & BL Lac & - \\
    FBQS J104309.0+240835 & FSRQ  & 0.56  & PKS 2320-035 & FSRQ  & 1.393 \\
    4C01.28 & Blaz.uncer & 0.89  & 2MASS J23385638+2124410 & BL Lac cand & 0.291 \\
    2XMM J110427.3+381231 & BL Lac & 0.031 &       &       &  \\
     \hline
    \end{tabular}
\end{table*}%
\end{center}
   \onecolumn
\begin{center}

\begin{longtable}[htbp!]{lrrrrrr}
  \caption[table]{List of objects with their RATAN-600 flux density values (Jy) and spectral indices ($\alpha$), measured between 2.3 and 7.7 GHz\label{fluxes}}\\
    \hline
    Name & $\alpha_{2.3-7.7GHz}$  & $F_{21.7GHz}$ & $F_{11.2GHz}$ & $F_{7.7GHz}$ & $F_{4.8GHz}$ & $F_{2.3GHz}$\\
		&                        & (Jy)                    & (Jy)                    & (Jy)                   & (Jy)                   & (Jy)                  \\
    \hline
    \endfirsthead
    \caption{continue}\\
    \hline
    Name & $\alpha_{2.3-7.7GHz}$  & $F_{21.7GHz}$ & $F_{11.2GHz}$ & $F_{7.7GHz}$ & $F_{4.8GHz}$ & $F_{2.3GHz}$\\
		&                        & (Jy)                    & (Jy)                    & (Jy)                   & (Jy)                   & (Jy)                  \\
    \hline
    \endhead
    \hline
    \endfoot
BZB J0022+0608	&	-0.10	&	$0.24\pm0.009$	&	$0.38\pm0.009$	&	$0.43\pm0.011$ &	$0.48\pm0.010$	&	$0.44\pm0.021$	\\
BZB J0035+1515	&	-0.40	&	$0.03\pm0.007$	&	$0.02\pm0.003$	&	$0.02\pm0.003$ &	$0.02\pm0.002$	&	$0.03\pm0.005$	\\
GC0039+23	    &	-0.29	&	$0.36\pm0.010$	&	$0.52\pm0.021$	&	$0.60\pm0.029$ &	$0.80\pm0.028$	&	$0.95\pm0.061$	\\
PKS 0047+023	&	-0.01	&	$0.14\pm0.020$	&	$0.19\pm0.007$	&	$0.20\pm0.007$ &	$0.23\pm0.005$	&	$0.18\pm0.009$	\\
FBQS J0050-0929	&	0.27	&	$1.42\pm0.022$	&	$1.32\pm0.038$	&	$1.26\pm0.027$ &	$1.06\pm0.017$	&	$0.78\pm0.029$	\\
PKS 0106+01	    &	0.23	&	$2.45\pm0.187$	&	$2.43\pm0.063$	&	$2.16\pm0.074$ &	$1.83\pm0.048$	&	$1.63\pm0.088$	\\
GC 0109+224	    &	0.24	&	$0.45\pm0.057$	&	$0.61\pm0.022$	&	$0.61\pm0.029$ &	$0.60\pm0.022$	&	$0.43\pm0.035$	\\
BZQ J0136+4751	& 0.43 &	$4.22\pm0.145$	&	$4.42\pm0.166$	&	$4.63\pm0.250$ &	$4.08\pm0.178$	&	$2.45\pm0.186$	\\
PKS 0139-09	    &	-0.19	&	$0.45\pm0.016$	&	$0.55\pm0.024$	&	$0.60\pm0.013$ &	$0.74\pm0.012$	&	$0.72\pm0.031$	\\
2MASX J01593439+1047052	&	-1.15	&	--	&	$0.02\pm0.003$	&	$0.02\pm0.003$ &	$0.04\pm0.003$	&	$0.08\pm0.010$	\\
4C +15.05	    &	-0.61	&	$0.74\pm0.033$	&	$1.09\pm0.042$	&	$1.36\pm0.061$ &	$1.89\pm0.058$	&	$2.85\pm0.376$	\\
2MASS J02171711+0837038	&	-0.18	&	$0.26\pm0.011$	&	$0.44\pm0.010$	&	$0.55\pm0.015$ &	$0.61\pm0.013$	&	$0.57\pm0.023$	\\
PKS 0215+015	&	0.33	&	$1.88\pm0.025$	&	$1.90\pm0.057$	&	$1.88\pm0.057$ &	$1.65\pm0.046$	&	$1.00\pm0.075$	\\
B2 0218+35	    &	-0.12	&	$1.32\pm0.018$	&	$1.50\pm0.120$	&	$1.71\pm0.070$ &	$1.59\pm0.075$	&	$1.78\pm0.103$	\\
3C 66A	        &	-0.32	&	$0.69\pm0.064$	&	$0.96\pm0.034$	&	$1.19\pm0.050$ &	$1.27\pm0.054$	&	$1.65\pm0.133$	\\
BZQ J0237+2848	&	-0.10	&	$1.89\pm0.038$	&	$2.71\pm0.089$	&	$2.86\pm0.123$ &	$3.06\pm0.108$	&	$3.16\pm0.176$	\\
PKS 0235+164	&	0.32	&	$4.44\pm0.053$	&	$5.38\pm0.133$	&	$5.54\pm0.164$ &	$5.16\pm0.127$	&	$2.90\pm0.113$	\\
2MASX J02503793+1712092	&	-0.57	&	$0.02\pm0.008$	&	$0.03\pm0.002$	&	$0.03\pm0.002$ &	$0.04\pm0.002$	&	$0.07\pm0.016$	\\
BZB J0303+4716	&	0.31	&	$0.99\pm0.057$	&	$1.05\pm0.039$	&	$1.17\pm0.057$ &	$0.95\pm0.059$	&	$0.71\pm0.094$	\\
BZB J0316+0904	&	-0.08	&	$0.07\pm0.018$	&	$0.07\pm0.005$	&	$0.07\pm0.006$ &	$0.08\pm0.003$	&	$0.08\pm0.012$	\\
NGC 1275	    &	0.83	&	$15.63\pm1.136$	&	$20.41\pm0.732$	&	$19.00\pm0.799$ &	$14.46\pm0.778$	&	$17.00\pm1.864$	\\
BZB J0319+1845	&	-0.83\footnote{\label{note1}measured between 4.8 and 7.7 GHz} &	$0.02\pm0.010$	&	$0.03\pm0.004$	&	$0.04\pm0.005$ &	$0.04\pm0.003$	&	--	\\
1H 0323+022	    &	-0.72	&	$0.02\pm0.007$	&	$0.02\pm0.003$	&	$0.03\pm0.003$ &	$0.05\pm0.002$	&	$0.08\pm0.014$	\\
PKS 0332-403	&	0.72	&	$2.12\pm0.041$	&	$1.77\pm0.108$	&	$1.46\pm0.102$ &	$0.86\pm0.059$	&	$0.45\pm0.040$	\\
NRAO 140	    &	-0.54	&	$1.48\pm0.129$	&	$1.16\pm0.045$	&	$1.13\pm0.054$ &	$1.54\pm0.073$	&	$2.50\pm0.195$	\\
PKS 0336-017	&	-0.03	&	$1.86\pm0.140$	&	$2.27\pm0.108$	&	$2.47\pm0.109$ &	$2.63\pm0.135$	&	$2.47\pm0.232$	\\
1H 0413+009	    &	-0.70	&	$0.03\pm0.005$	&	$0.04\pm0.003$	&	$0.05\pm0.004$ &	$0.07\pm0.002$	&	$0.11\pm0.008$	\\
PKS 0420+022	&	-0.03	&	$0.64\pm0.044$	&	$0.97\pm0.021$	&	$1.16\pm0.028$ &	$1.23\pm0.023$	&	$0.94\pm0.029$	\\
PKS 0420-01	    &	0.54	&	$5.10\pm0.477$	&	$4.75\pm0.692$	&	$4.18\pm0.607$ &	$3.22\pm0.232$	&	$2.09\pm0.212$	\\
PKS 0422+004	&	0.03	&	$0.60\pm0.015$	&	$0.68\pm0.014$	&	$0.65\pm0.016$ &	$0.66\pm0.012$	&	$0.55\pm0.024$	\\
PKS 0446+11	    &	-0.11	&	$0.72\pm0.074$	&	$0.66\pm0.053$	&	$0.67\pm0.040$ &	$0.80\pm0.044$	&	$0.88\pm0.141$	\\
PKS 0454-234	&	0.06	&	$2.08\pm0.022$	&	$2.15\pm0.072$	&	$2.23\pm0.074$ &	$2.26\pm0.087$	&	$2.13\pm0.088$	\\
4C-02.19	    &	0.07	&	$1.03\pm0.085$	&	$0.89\pm0.062$	&	$0.90\pm0.044$ &	$0.93\pm0.025$	&	$0.85\pm0.093$	\\
BZQ J0505+0459  &	-0.84	&	$0.03\pm0.005$	&	$0.06\pm0.003$	&	$0.08\pm0.005$ &	$0.12\pm0.003$	&	$0.19\pm0.014$	\\
2MASS J05075617+6737242	&	0.12\textsuperscript{\ref{note1}}	&	$0.05\pm0.015$	&	$0.04\pm0.008$	&	$0.04\pm0.010$ &	$0.04\pm0.006$	&	--	\\
BZB J0509+0541	&	-0.02	&	$0.45\pm0.020$	&	$0.67\pm0.015$	&	$0.69\pm0.018$ &	$0.70\pm0.015$	&	$0.64\pm0.022$	\\
PKS 0507+17	    &	0.38	&	$1.30\pm0.019$	&	$1.17\pm0.043$	&	$0.94\pm0.045$ &	$0.71\pm0.050$	&	$0.48\pm0.064$	\\
PKS 0539-057	&	-0.02	&	$1.08\pm0.027$	&	$1.18\pm0.107$	&	$1.30\pm0.061$ &	$1.38\pm0.035$	&	$1.22\pm0.114$	\\
OH-10	        &	-0.16	&	$1.49\pm0.146$	&	--	&	$0.98\pm0.072$ &	$1.11\pm0.028$	&	$1.33\pm0.136$	\\
2MASS J06251826+4440014	&	-0.02	&	$0.13\pm0.014$	&	$0.21\pm0.010$	&	$0.22\pm0.013$ &	$0.20\pm0.014$	&	$0.18\pm0.030$	\\
2MASX J07103005+5908202	&	-0.88	&	$0.12\pm0.036$	&	$0.08\pm0.011$	&	$0.07\pm0.006$ &	$0.10\pm0.009$	&	$0.19\pm0.016$	\\
PKS 0723-008	&	0.25	&	$2.26\pm0.131$	&	$2.14\pm0.062$	&	$2.18\pm0.065$ &	$2.34\pm0.052$	&	$1.52\pm0.084$	\\
PKS 0735+17	    &	-0.31	&	$0.49\pm0.011$	&	$0.75\pm0.019$	&	$0.84\pm0.026$ &	$0.96\pm0.024$	&	$1.11\pm0.045$	\\
PKS 0736+01	    &	-0.08	&	$1.32\pm0.090$	&	$1.43\pm0.045$	&	$1.48\pm0.062$ &	$1.59\pm0.062$	&	$1.58\pm0.260$	\\
PKS 0748+126	&	0.36	&	$3.65\pm0.112$	&	$4.18\pm0.112$	&	$4.13\pm0.164$ &	$3.68\pm0.206$	&	$2.49\pm0.282$	\\
PKS 0754+100	&	0.02	&	$0.93\pm0.015$	&	$1.02\pm0.023$	&	$1.03\pm0.027$ &	$0.98\pm0.021$	&	$0.92\pm0.034$	\\
PKS 0805-07	    &	0.86	&	$1.54\pm0.119$	&	$0.97\pm0.143$	&	$0.85\pm0.039$ &	$0.39\pm0.028$	&	$0.59\pm0.076$	\\
PKS 0808+019	&	0.77	&	$1.15\pm0.018$	&	$1.04\pm0.021$	&	$0.83\pm0.020$ &	$0.49\pm0.009$	&	$0.31\pm0.014$	\\
B3 0814+425	    &	0.25	&	$1.32\pm0.184$	&	$1.70\pm0.049$	&	$1.81\pm0.087$ &	$1.52\pm0.080$	&	$1.20\pm0.135$	\\
PKS 0823+033	&	0.38	&	$1.87\pm0.028$	&	$1.95\pm0.041$	&	$1.74\pm0.042$ &	$1.38\pm0.026$	&	$1.07\pm0.049$	\\
BZQ J0830+2410	&	0.37	&	$1.09\pm0.173$	&	$1.48\pm0.062$	&	$1.46\pm0.059$ &	$1.29\pm0.121$	&	$0.85\pm0.163$	\\
PKS 0829+046	&	-0.08	&	$0.55\pm0.038$	&	$0.59\pm0.021$	&	$0.59\pm0.018$ &	$0.59\pm0.030$	&	$0.68\pm0.199$	\\
2EG J0852-1237	&	0.41	&	$0.83\pm0.048$	&	$0.73\pm0.034$	&	$0.64\pm0.022$ &	$0.45\pm0.046$	&	$0.38\pm0.074$	\\
PKS 0851+202	&	0.63	&	$3.81\pm0.165$	&	$3.44\pm0.111$	&	$2.65\pm0.114$ &	$1.95\pm0.073$	&	$1.24\pm0.090$	\\
BZQ J0920+4441	&	0.39	&	$2.20\pm0.146$	&	$1.81\pm0.058$	&	$1.50\pm0.053$ &	$1.08\pm0.045$	&	$0.99\pm0.052$	\\
2MASS J09303759+4950256	&	-0.71	&	--	&	$0.06\pm0.014$	&	$0.06\pm0.009$ &	$0.07\pm0.013$	&	$0.16\pm0.019$	\\
OK 290	        &	-0.20	&	$0.62\pm0.041$	&	$0.70\pm0.027$	&	$0.71\pm0.035$ &	$0.83\pm0.039$	&	$0.97\pm0.133$	\\
2MASS J10121335+0630569	&	-0.49	&	$0.13\pm0.012$	&	$0.20\pm0.006$	&	$0.23\pm0.008$ &	$0.32\pm0.011$	&	$0.44\pm0.021$	\\
SDSS J101603.13+051302.3	&	0.11	&	$0.68\pm0.012$	&	$0.71\pm0.020$	&	$0.68\pm0.029$ &	$0.69\pm0.028$	&	$0.62\pm0.068$	\\
FBQS J104309.0+240835	&	0.08	&	$0.41\pm0.009$	&	$0.60\pm0.017$	&	$0.67\pm0.023$ &	$0.74\pm0.021$	&	$0.57\pm0.025$	\\
4C01.28	        &	0.20	&	$4.52\pm0.147$	&	$4.09\pm0.102$	&	$3.57\pm0.104$ &	$3.20\pm0.084$	&	$2.98\pm0.322$	\\
2XMM J110427.3+381231	&	-0.36	&	$0.29\pm0.010$	&	$0.41\pm0.013$	&	$0.48\pm0.019$ &	$0.53\pm0.019$	&	$0.68\pm0.034$	\\
PKS 1118-05	    &	-0.31	&	$0.33\pm0.016$	&	$0.45\pm0.022$	&	$0.55\pm0.012$ &	$0.75\pm0.014$	&	$0.87\pm0.033$	\\
FBQS J115019.2+241753	&	0.04	&	$0.56\pm0.015$	&	$0.73\pm0.020$	&	$0.78\pm0.027$ &	$0.82\pm0.024$	&	$0.71\pm0.031$	\\
4C 29.45	    &	0.17	&	$1.92\pm0.110$	&	$2.65\pm0.084$	&	$2.68\pm0.115$ &	$2.71\pm0.092$	&	$2.07\pm0.178$	\\
EXO1218.8+3027	&	-0.57	&	$0.02\pm0.005$	&	$0.03\pm0.003$	&	$0.04\pm0.004$ &	$0.05\pm0.002$	&	$0.08\pm0.013$	\\
ON 231	        &	0.09	&	$0.33\pm0.078$	&	$0.43\pm0.025$	&	$0.44\pm0.025$ &	$0.49\pm0.017$	&	$0.38\pm0.067$	\\
PKS 1219+04	    &	0.36	&	$1.04\pm0.105$	&	$1.06\pm0.058$	&	$0.82\pm0.044$ &	$0.64\pm0.013$	&	$0.56\pm0.076$	\\
PKS 1222+21	    &	-0.33	&	$0.76\pm0.120$	&	$1.13\pm0.041$	&	$1.18\pm0.056$ &	$1.31\pm0.067$	&	$1.80\pm0.182$	\\
3C 273	        &	-0.27	&	$21.63\pm1.771$	&	$28.48\pm0.660$	&	$32.85\pm1.606$ &	$39.02\pm1.006$	&	$44.79\pm3.988$	\\
FBQS J123014.0+251807	&	-0.24	&	$0.11\pm0.008$	&	$0.16\pm0.005$	&	$0.18\pm0.007$ &	$0.23\pm0.007$	&	$0.24\pm0.015$	\\
3C 279	        &	0.26	&	$13.66\pm0.615$	&	$12.67\pm0.447$	&	$11.85\pm0.305$ &	$10.84\pm0.240$	&	$8.51\pm0.341$	\\
1WGA J1310.4+3220	&	0.32	&	$1.79\pm0.063$	&	$1.60\pm0.054$	&	$1.34\pm0.059$ &	$1.11\pm0.041$	&	$0.93\pm0.173$	\\
GB6 J1327+2210	&	-0.32	&	$0.52\pm0.070$	&	$0.77\pm0.035$	&	$0.91\pm0.043$ &	$1.16\pm0.041$	&	$1.31\pm0.162$	\\
PKS 1335-127	&	0.50	&	$5.77\pm0.057$	&	$4.97\pm0.143$	&	$4.83\pm0.280$ &	$3.91\pm0.163$	&	$2.12\pm0.104$	\\
FIRST J134105.1+395945	&	-0.98	&	--	&	$0.02\pm0.004$	&	$0.02\pm0.004$ &	$0.04\pm0.004$	&	$0.09\pm0.021$	\\
CGRaBS J1357+7643	&	-0.05	&	$0.55\pm0.016$	&	$0.76\pm0.031$	&	$0.82\pm0.032$ &	$0.87\pm0.042$	&	$0.78\pm0.061$	\\
PKS 1406-076	&	0.15	&	$0.77\pm0.074$	&	$0.75\pm0.071$	&	$0.78\pm0.028$ &	$0.71\pm0.018$	&	$0.61\pm0.082$	\\
FBQS J142700.4+234800	&	-0.23	&	$0.18\pm0.011$	&	$0.28\pm0.008$	&	$0.31\pm0.011$ &	$0.37\pm0.011$	&	$0.38\pm0.021$	\\
2MASS J14424821+1200402	&	-0.49	&	$0.02\pm0.006$	&	$0.03\pm0.002$	&	$0.03\pm0.002$ &	$0.04\pm0.002$	&	$0.06\pm0.010$	\\
PKS 1502+106	&	0.48	&	$3.32\pm0.103$	&	$2.79\pm0.081$	&	$2.29\pm0.079$ &	$1.65\pm0.046$	&	$1.29\pm0.092$	\\
PKS 1502+036	&	-0.01	&	$0.53\pm0.018$	&	$0.60\pm0.032$	&	$0.62\pm0.028$ &	$0.63\pm0.031$	&	$0.57\pm0.023$	\\
PKS 1510-08	    &	0.14	&	$2.48\pm0.204$	&	$2.16\pm0.244$	&	$1.80\pm0.069$ &	$1.83\pm0.047$	&	$1.84\pm0.284$	\\
PKS 1514+197	&	0.75	&	$1.22\pm0.017$	&	$1.18\pm0.031$	&	$0.99\pm0.031$ &	$0.74\pm0.019$	&	$0.37\pm0.018$	\\
1RXS J152239.7-273025	&	0.10	&	$0.65\pm0.022$	&	$0.85\pm0.030$	&	$0.99\pm0.062$ &	$1.12\pm0.066$	&	$0.94\pm0.133$	\\
4C +05.64	    &	-0.06	&	$2.59\pm0.094$	&	$3.00\pm0.072$	&	$3.20\pm0.100$ &	$3.54\pm0.096$	&	$3.34\pm0.186$	\\
BZB J1555+1111  &	-0.34	&	$0.12\pm0.010$	&	$0.16\pm0.005$	&	$0.18\pm0.007$ &	$0.19\pm0.005$	&	$0.26\pm0.016$	\\
SDSS J160706.23+155136.8	&	-0.27	&	$0.29\pm0.012$	&	$0.38\pm0.010$	&	$0.39\pm0.012$ &	$0.44\pm0.011$	&	$0.52\pm0.022$	\\
4C +10.45	    &	-0.25	&	$0.98\pm0.037$	&	$1.25\pm0.043$	&	$1.32\pm0.052$ &	$1.45\pm0.042$	&	$1.81\pm0.154$	\\
BZQ J1613+3412	&	-0.25	&	$1.90\pm0.095$	&	$3.12\pm0.112$	&	$3.45\pm0.154$ &	$3.84\pm0.146$	&	$4.58\pm0.685$	\\
4C +38.41	    &	0.12	&	$2.29\pm0.117$	&	$2.90\pm0.091$	&	$2.95\pm0.124$ &	$2.94\pm0.115$	&	$2.47\pm0.144$	\\
3C 345	        &	0.38	&	$6.71\pm0.245$	&	$7.25\pm0.234$	&	$6.54\pm0.273$ &	$5.90\pm0.241$	&	$6.99\pm0.652$	\\
MRK 0501	    &	0.07	&	$0.94\pm0.042$	&	$1.31\pm0.040$	&	$1.43\pm0.059$ &	$1.46\pm0.057$	&	$1.49\pm0.102$	\\
PKS 1717+177	&	0.00	&	$0.43\pm0.015$	&	$0.58\pm0.015$	&	$0.62\pm0.020$ &	$0.60\pm0.016$	&	$0.55\pm0.025$	\\
2MASS J17250434+1152155	&	-0.55	&	$0.04\pm0.008$	&	$0.05\pm0.004$	&	$0.05\pm0.005$ &	$0.06\pm0.003$	&	$0.10\pm0.012$	\\
NRAO 530	    &	0.12	&	$4.57\pm0.165$	&	$4.76\pm0.104$	&	$4.45\pm0.125$ &	$4.03\pm0.078$	&	$3.85\pm0.179$	\\
1ES 1741+196	&	-0.64	&	$0.10\pm0.005$	&	$0.14\pm0.005$	&	$0.17\pm0.007$ &	$0.21\pm0.006$	&	$0.38\pm0.019$	\\
OT 081	        &	0.59	&	$5.58\pm0.065$	&	$5.35\pm0.120$	&	$4.42\pm0.116$ &	$3.16\pm0.068$	&	$1.60\pm0.063$	\\
BZB J1756+5522  &	-0.95	&	--	&	$0.06\pm0.009$	&	$0.05\pm0.008$ &	$0.06\pm0.008$	&	$0.22\pm0.040$	\\
BZB J1800+7828	&	0.18	&	$3.32\pm0.038$	&	$2.85\pm0.102$	&	$2.79\pm0.090$ &	$2.71\pm0.112$	&	$1.97\pm0.107$	\\
3C 380.0	    &	-0.63	&	$2.97\pm0.115$	&	$3.44\pm0.094$	&	$4.04\pm0.130$ &	$5.32\pm0.210$	&	$9.03\pm0.643$	\\
BZQ J1852+4855	&	-0.15	&	$0.25\pm0.107$	&	$0.22\pm0.033$	&	$0.21\pm0.031$ &	$0.23\pm0.014$	&	$0.27\pm0.022$	\\
PKS 1954-388	&	-0.10	&	$1.20\pm0.025$	&	$1.92\pm0.089$	&	$2.19\pm0.085$ &	$2.37\pm0.103$	&	$2.14\pm0.115$	\\
BZB J2005+7752	&	0.11	&	$0.88\pm0.018$	&	$0.89\pm0.043$	&	$0.89\pm0.030$ &	$0.86\pm0.037$	&	$0.78\pm0.046$	\\
PKS 2012-017	&	-0.32	&	$0.30\pm0.012$	&	$0.44\pm0.010$	&	$0.48\pm0.012$ &	$0.58\pm0.011$	&	$0.68\pm0.022$	\\
BZQ J2035+1056	&	-0.34	&	$0.32\pm0.016$	&	$0.38\pm0.010$	&	$0.42\pm0.013$ &	$0.57\pm0.013$	&	$0.69\pm0.026$	\\
PKS 2047+039	&	-0.17	&	$0.42\pm0.008$	&	$0.64\pm0.014$	&	$0.69\pm0.017$ &	$0.73\pm0.015$	&	$0.77\pm0.026$	\\
PKS 2131-021	&	-0.08	&	$1.91\pm0.021$	&	$2.22\pm0.043$	&	$2.29\pm0.051$ &	$2.51\pm0.043$	&	$2.39\pm0.070$	\\
BZQ J2143+1743	&	0.29	&	$0.74\pm0.080$	&	$1.02\pm0.035$	&	$1.04\pm0.052$ &	$0.96\pm0.030$	&	$0.67\pm0.093$	\\
4C 06.69	    &	0.30	&	$4.52\pm0.567$	&	$5.52\pm0.178$	&	$5.72\pm0.169$ &	$5.59\pm0.154$	&	$3.65\pm0.181$	\\
PKS 2149+173	&	-0.28	&	$0.35\pm0.016$	&	$0.54\pm0.015$	&	$0.60\pm0.019$ &	$0.68\pm0.018$	&	$0.77\pm0.036$	\\
BL Lac	        &	0.28	&	$2.88\pm0.330$	&	$3.72\pm0.314$	&	$3.85\pm0.242$ &	$3.06\pm0.140$	&	$2.46\pm0.149$	\\
PKS 2201+171	&	0.44	&	$0.62\pm0.031$	&	$0.80\pm0.026$	&	$0.73\pm0.042$ &	$0.68\pm0.041$	&	$0.87\pm0.067$	\\
PKS 2209+236	&	0.03	&	$0.68\pm0.016$	&	$0.91\pm0.034$	&	$0.98\pm0.042$ &	$1.07\pm0.053$	&	$0.79\pm0.095$	\\
3C 446	        &	0.18	&	$6.59\pm0.351$	&	$7.87\pm0.395$	&	$7.53\pm0.228$ &	$6.85\pm0.159$	&	$5.89\pm0.196$	\\
BZQ J2229-0832  &	0.35	&	$2.31\pm0.168$	&	$2.41\pm0.175$	&	$2.42\pm0.085$ &	$2.40\pm0.059$	&	$1.48\pm0.135$	\\
4C -11.69	    &	-0.12	&	$4.10\pm0.402$	&	$4.25\pm0.275$	&	$4.27\pm0.152$ &	$4.58\pm0.156$	&	$5.32\pm0.296$	\\
3C 454.3	    &	-0.03	&	$14.11\pm0.159$	&	$11.79\pm0.893$	&	$10.79\pm0.350$ &	$10.18\pm0.476$	&	$12.44\pm0.546$	\\
PKS 2254-204	&	0.19	&	$1.08\pm0.030$	&	$1.00\pm0.025$	&	$1.06\pm0.042$ &	$1.04\pm0.091$	&	$0.75\pm0.090$	\\
PKS 2255-282	&	0.67	&	$3.58\pm0.037$	&	$2.80\pm0.096$	&	$2.75\pm0.108$ &	$2.21\pm0.143$	&	$1.09\pm0.052$	\\
BZB J2304+3705	&	-0.71	&	$0.02\pm0.003$	&	$0.02\pm0.002$	&	$0.01\pm0.002$ &	$0.02\pm0.001$	&	$0.04\pm0.005$	\\
PKS 2320-035	&	0.07	&	$0.92\pm0.051$	&	$1.01\pm0.057$	&	$1.04\pm0.038$ &	$1.05\pm0.032$	&	$0.93\pm0.061$	\\
2MASS J23385638+2124410	&	-0.73	&	$0.02\pm0.004$	&	$0.02\pm0.002$	&	$0.02\pm0.002$ &	$0.04\pm0.002$	&	$0.06\pm0.012$	\\
\hline
\end{longtable}%
\end{center}
\twocolumn 
\begin{table*}
\caption{Pearson's $r$ for broadband gamma-ray emission (photon fluxes at five bands 0.1 -- 100 GeV) and radio emission (flux densities at five frequencies 2.3 -- 21.7 GHz). For each frequency the number of sources (N) and typical confidence level values for correlation in this bands (CL) are presented. In Figure \ref{Cor-Energy} we show the correlation coefficient distribution across the five energy bands and five frequencies, reported in this table.}
\label{table_corr}

\begin{tabular}{|c|c|c|c|c|c|c|}
\hline
Source class & N & 0.1--0.3 GeV & 0.3--1 GeV & 1--3 GeV & 3--10 GeV & 10--100 GeV \\
\hline
\multicolumn{7}{|c|}{21.7 GHz}\\
\hline
BL Lac & 50 & $+0.61$ & $+0.53$ & $+0.43$ & $+0.25$ & $-0.03$ \\
FSRQ & 56 &  $+0.57$ & $+0.46$ & $+0.39$ & $+0.26$ & $+0.23$ \\ 
\hline
\multicolumn{7}{|c|}{11.2 GHz}\\
\hline
BL Lac & 53 & $+0.69$ & $+0.62$ & $+0.51$ & $+0.32$ & $+0.01$ \\
FSRQ & 55 &  $+0.47$ & $+0.36$ & $+0.28$ & $+0.16$ & $+0.16$ \\ 
\hline
\multicolumn{7}{|c|}{7.7 GHz}\\
\hline
BL Lac & 53 & $+0.73$ & $+0.67$ & $+0.56$ & $+0.36$ & $+0.04$ \\
FSRQ & 56 &  $+0.42$ & $+0.31$ & $+0.23$ & $+0.11$ & $+0.11$ \\ 
\hline
\multicolumn{7}{|c|}{4.8 GHz}\\
\hline
BL Lac & 53 & $+0.74$ & $+0.69$ & $+0.59$ & $+0.39$ & $+0.06$ \\
FSRQ & 56 &  $+0.39$ & $+0.26$ & $+0.18$ & $+0.07$ & $+0.07$ \\ 
\hline
\multicolumn{7}{|c|}{2.3 GHz}\\
\hline
BL Lac & 51 & $+0.64$ & $+0.64$ & $+0.55$ & $+0.39$ & $+0.14$ \\ 
FSRQ & 56 & $+0.39$ & $+0.26$ & $+0.18$ & $+0.05$ & $+0.03$ \\ 
\hline
\hline
\multicolumn{7}{|c|}{CL}\\
\hline
BL Lac & & 99\% & 99\% & 99\% & 99\% & $<$90\% \\ 
FSRQ & &  99\% & 95\% & 90\% & $<$90\% & $<$90\% \\ 
\hline
\end{tabular}
\end{table*}

\begin{table*}
\caption{Pearson's $r$ for the radio flux density and gamma-ray photon fluxes with their significances ($sig$), calculated from data randomization analysis, for sources with known redshift (their number marked with N).}
\label{table_sig}

\begin{tabular}{|c|c|c|c|c|c|c|c|c|c|c|c|}
\hline
$E$ (GeV) & &\multicolumn{2}{|c|}{0.1--0.3} &\multicolumn{2}{|c|}{0.3--1} &\multicolumn{2}{|c|}{1--3} &\multicolumn{2}{|c|}{3--10} &\multicolumn{2}{|c|}{10--100} \\
\hline
Source class & N & $r$ & $sig$ & $r$ & $sig$ & $r$ & $sig$ & $r$ & $sig$ & $r$ & $sig$\\
\hline
\multicolumn{12}{|c|}{21.7 GHz}\\
\hline
BL Lac & 45 & $+0.61$ & $5.29\times 10^{-7}$ & $+0.53$ & $3.76\times 10^{-5}$ & $+0.42$ & $1.13\times 10^{-3}$ & $+0.24$ & $0.055$ & $-0.05$ & $0.629$  \\
FSRQ & 55 &  $+0.56$ & $6.26\times 10^{-7}$ & $+0.46$ & $9.79\times 10^{-5}$ & $+0.39$ & $1.13\times 10^{-3}$ & $+0.26$ & $0.023$ & $+0.23$ & $0.043$  \\ 
\hline
\multicolumn{12}{|c|}{11.2 GHz}\\
\hline
BL Lac & 48 & $+0.70$ & $4.14\times 10^{-7}$ & $+0.62$ & $1.32\times 10^{-6}$ & $+0.51$ & $8.21\times 10^{-5}$ & $+0.31$ & $0.016$ & $-0.003$ & $0.934$  \\
FSRQ & 54 &  $+0.47$ & $7.74\times 10^{-5}$ & $+0.36$ & $2.47\times 10^{-3}$ & $+0.28$ & $0.016$ & $+0.16$ & $0.134$ & $+0.15$ & $0.15$  \\ 
\hline
\multicolumn{12}{|c|}{7.7 GHz}\\
\hline
BL Lac & 48 & $+0.74$ & $3.36\times 10^{-7}$ & $+0.67$ & $4.28\times 10^{-7}$ & $+0.56$ & $1.38\times 10^{-5}$ & $+0.35$ & $6.54\times 10^{-3}$ & $+0.03$ & $0.796$  \\
FSRQ & 55 &  $+0.42$ & $4.21\times 10^{-4}$ & $+0.30$ & $9.68\times 10^{-3}$ & $+0.23$ & $0.048$ & $+0.11$ & $0.278$ & $+0.11$ & $0.288$  \\ 
\hline
\multicolumn{12}{|c|}{4.8 GHz}\\
\hline
BL Lac & 48 & $+0.75$ & $3.24\times 10^{-7}$ & $+0.69$ & $4.06\times 10^{-7}$ & $+0.58$ & $5.44\times 10^{-6}$ & $+0.38$ & $3.14\times 10^{-3}$ & $+0.05$ & $0.632$  \\
FSRQ & 55 &  $+0.39$ & $1.28\times 10^{-3}$ & $+0.26$ & $0.025$ & $+0.18$ & $0.104$ & $+0.07$ & $0.478$ & $+0.07$ & $0.480$  \\ 
\hline
\multicolumn{12}{|c|}{2.3 GHz}\\
\hline
BL Lac & 46 & $+0.63$ & $4.87\times 10^{-7}$ & $+0.63$ & $4.94\times 10^{-7}$ & $+0.54$ & $2.35\times 10^{-5}$ & $+0.38$ & $2.87\times 10^{-3}$ & $+0.13$ & $0.28$  \\ 
FSRQ & 55 & $+0.39$ & $1.08\times 10^{-3}$ & $+0.26$ & $0.025$ & $+0.18$ & $0.110$ & $+0.05$ & $0.562$ & $+0.04$ & $0.695$  \\ 
\hline
\end{tabular}
\end{table*}




\end{document}